\documentclass[%
 aip,
 amsmath,amssymb,
 longbibliography,
 preprint,%
 superscriptaddress,
 sort,
 compress,
]{revtex4-2}
\usepackage{graphicx}
\usepackage{dcolumn}
\usepackage{bm}
\usepackage[mathlines]{lineno}
\usepackage[utf8]{inputenc}
\usepackage[T1]{fontenc}
\usepackage{mathptmx}
\usepackage[varg]{newtxmath} %
\usepackage{etoolbox}
\usepackage{caption}
\usepackage{subcaption}
\usepackage{booktabs}
\usepackage{multirow}

\usepackage{hyperref}
\hypersetup{
    colorlinks=true,       
    linkcolor=blue,        
    citecolor=blue,        
    urlcolor=blue         
}
\usepackage{xcolor}

\renewcommand{\selectlanguage}[1]{}

\begin{document}

\title{Simulation of Weakly Ionized Hypersonic Flows with Reactive Species Weighting Scheme in the Direct Simulation Monte Carlo Method}

\author{Takato Morimoto}
\email[Author to whom correspondence should be addressed: ]{morimoto.takato.c8@s.mail.nagoya-u.ac.jp}
\affiliation{Department of Aerospace Engineering, Nagoya University, Nagoya, Aichi 464-8603, Japan}
\author{Virgile Charton}
\affiliation{Department of Aerospace Engineering, Nagoya University, Nagoya, Aichi 464-8603, Japan}
\author{Eiichiro Yamaoka}
\affiliation{Department of Aerospace Engineering, Nagoya University, Nagoya, Aichi 464-8603, Japan}
\author{Kiyoshi Kinefuchi}
\affiliation{Department of Aerospace Engineering, Nagoya University, Nagoya, Aichi 464-8603, Japan}

\begin{abstract}
The conservative species weighting scheme enables the direct simulation Monte Carlo (DSMC) method to resolve trace species such as electrons and ions in weakly ionized flows. The present study extends this scheme to reacting flows, utilizing a split-merge concept for particles with species-dependent weights. Reactive collisions between differently weighted species are processed according to their respective weights, introducing the production of an integer number of product particles and a probabilistic approach applied to both reactant and product particles to satisfy mass conservation. The extended scheme was validated in single-cell dissociation and exchange reaction test cases, verifying mass and total energy conservation and the mole-fraction histories against conventional DSMC. This extension provides the electron number density required for predicting the onset of communication blackout at high altitudes during hypersonic reentry. The method was applied to the Orbital Re-entry Experiment (OREX) flight conditions from 105.0 km to 92.8 km altitude. Translational and internal temperatures showed good agreement with conventional DSMC results, validating the macroscopic flow-field computation. The scheme enabled continuous spatial distributions of trace electron number density, consistent with the communication blackout onset observed in the flight data. The computed ion current reproduced the overall trend of the in-flight electrostatic probe data. The influences of the ion recombination at the vehicle surface and of the uncertainty of the $\mathrm{N}+\mathrm{O}$ associative-ionization rate were quantified, and the range of the computed currents encompassed the available measurements. The 105.0 km case extended the prediction beyond available measurements, consistent with the overall trend.
\end{abstract}
\maketitle

\section{Introduction}
\label{sec:intro}

During vehicle reentry at hypersonic speed into the atmosphere, aerodynamic heating causes air dissociation and ionization in the high temperature shock layer. This process generates a plasma layer that can interfere with electromagnetic waves, leading to a loss of communication known as radio-frequency (RF) blackout. 
Under the assumption that the electron--heavy particle collision frequency is sufficiently low compared with the signal frequency, blackout onset is determined by whether the electron number density $n_{\mathrm{e}}$ exceeds a critical value $n_{\mathrm{e},\mathrm{crit}}$, defined by the condition that the plasma frequency equals the signal frequency $f$:
\begin{equation}
    n_{\mathrm{e},\mathrm{crit}} = \frac{4\pi^2 \varepsilon_0 m_{\mathrm{e}}}{e^2} f^2,
    \label{eq:ne_crit}
\end{equation}
where $\varepsilon_0$ is the permittivity of free space, $m_{\mathrm{e}}$ is the electron mass, and $e$ is the elementary charge. Since this interference can be mitigated by either reducing the electron number density or increasing the signal frequency, mitigation strategies for the blackout have been actively studied, including electromagnetic field manipulation, gas injection, and the use of high-frequency terahertz waves~\cite{cheng_electromagnetic_2017, miyashita_communication_2025, sun_open_2024, maicke_mitigation_2026}.
Accurate prediction of the electron number density distribution around the vehicle is therefore essential both for assessing blackout risks and for guiding the design of such countermeasures.
While high-density, low-altitude regimes are frequently analyzed using continuum approaches based on the Navier--Stokes equations, including analyses of the plasma flow around reentry vehicles~\cite{candler_computation_1991, farbar_numerical_2013, jung_numerical_2016}, blackout phenomena also occur at high altitudes where the flow can be rarefied.

In these high-altitude regimes, the mean free path of gas molecules is not negligible compared to the characteristic length scales of the vehicle. When the Knudsen number is sufficiently large, collision frequencies become too low to maintain local thermodynamic equilibrium. Under such conditions, continuum assumptions break down, and macroscopic Navier--Stokes solvers are not suitable. Instead, microscopic kinetic approaches based on the Boltzmann equation are required to capture the rarefied gas characteristics. Among these, the direct simulation Monte Carlo (DSMC) method~\cite{bird_molecular_1976} has established itself as the standard numerical technique for simulating rarefied gas flows.

The DSMC method solves flow fields by tracking the motion and collisions of representative numerical particles. In the standard DSMC formulations, the ratio of real molecules to simulated particles $F_N$ (the global weighting factor) is constant across the domain. This poses a challenge for weakly ionized hypersonic flows, where the number density of charged species (electrons and ions) can be several orders of magnitude lower than that of neutral species. To resolve the statistical behavior of these trace species, an enormous number of particles for the dominant neutral species would be required, leading to prohibitive computational costs and memory requirements. Furthermore, since the cell size in DSMC should be smaller than the mean free path to accurately select collisions, obtaining detailed electron density fields becomes computationally infeasible in standard frameworks. 

To overcome these computational limitations, a species weighting scheme (SWS) has been developed, and the present study extends the conservative SWS to chemically reactive flows. In the SWS, simulated particles are assigned species-dependent numerical weights $w_i$, representing different numbers of real molecules, which allows trace species to be simulated efficiently without over-populating the domain with major species. However, such schemes require special treatments to satisfy the conservation of mass, momentum, and energy in collisions between differently weighted particles.
Bird~\cite{bird_molecular_1994} initially proposed a non-conservative scheme where collisions between particles of different weights are handled probabilistically, but this method does not explicitly conserve momentum and energy in each collision. Boyd~\cite{boyd_conservative_1996} subsequently developed a conservative approach using a split-merge method to ensure momentum and energy conservation in such collisions. In non-reactive systems, He and Zhao~\cite{he_conservative_2016} applied this conservative SWS in DSMC coupled with the lattice-Boltzmann method for gas-solid flows. Charton et al.~\cite{charton_species_2025} further investigated conservative SWS for jet flows, clarifying appropriate pair selection procedures that account for species weights.
Extending the SWS to chemically reactive flows introduces additional complexity, as the split-merge procedure cannot be directly applied when reactants and products possess different weights. For chemical vapor deposition processes, Wu et al.~\cite{wu_assessment_2003} and Sakiyama et al.~\cite{sakiyama_full_2001} employed conservative species weighting approaches by assigning weights to individual simulated particles, rather than to species. However, this leads to a significant increase in the total number of particles during computation, making computational load control difficult. Petkow et al.~\cite{petkow_treatment_2012} proposed a split-merge based reaction model for high temperature gases, but no practical application results were presented. Fang et al.~\cite{fang_dsmc_2020} employed Bird's non-conservative scheme for hypersonic plasma prediction. Shevyrin and Bondar~\cite{shevyrin_application_2024} also applied a different probabilistic weighting scheme to weakly ionized reentry flows without a split-merge energy conservation mechanism. The present study extends the work of Charton et al.~\cite{charton_species_2025} to reactive flows, introducing a probabilistic mass conservation approach for chemical reactions within the conservative SWS framework. This method maintains the assigned weight for each species throughout the simulation, thereby keeping the computational load controllable.

We apply this approach to obtain and investigate the plasma flow around the Orbital Re-entry Experiment (OREX) vehicle, focusing on the flight segment from 105.0~km~--~92.8~km altitude to predict blackout onset. The OREX flight, conducted in 1994, provided valuable flight data, including electrostatic probe measurements of electron density and observations of communication blackout~\cite{orex_1994}. Among the phenomena observed in relation to the blackout, there were discrepancies in the blackout timing reported by different base stations. The OREX vehicle was equipped with telemetry antennas on the aft body. Because the vehicle was descending toward the ground stations, the communication paths from the aft-mounted antennas were affected by the plasma layer developing on the forebody side. The observed discrepancies in blackout timing between stations therefore suggest the development of a plasma layer at high altitudes on the stagnation side of the vehicle, which the present study aims to characterize. While previous DSMC studies on OREX have focused primarily on aerodynamic heating~\cite{moss_dsmc_1996, schouler_survey_2020}, this work demonstrates the capability of the proposed SWS to obtain high-altitude electron number density profiles that were previously computationally inaccessible. The resulting plasma profiles can then be used for electromagnetic wave propagation analysis during hypersonic flight, following the approach of Morimoto and Kinefuchi~\cite{morimoto_effect_2025}, thereby enabling accurate prediction of communication blackout onset. The sensitivity of the predicted plasma quantities to the uncertain ionization rates and to the ion recombination at the vehicle surface is also quantified. Details of the OREX configuration and flight conditions are described in Sec.~\ref{sec:case_model}.

This paper is structured as follows. Section~\ref{sec:numerical_method} details the DSMC method and the implementation of the conservative species weighting scheme, including the handling of reactive collisions. Section~\ref{sec:case_model} outlines the physical models and the OREX analysis conditions. Section~\ref{sec:results_discussion} presents the computational results and discusses the flow-field characteristics and method validity through comparison of electron number density with flight measurement data. Finally, Section~\ref{sec:conclusion} concludes the paper.

\section{Numerical Methods}
\label{sec:numerical_method}

The DSMC method calculates flow field behavior by representing real flow molecules with simulated particles and computing their trajectories and collisions at each time step. Specifically, the motion of particles is firstly calculated according to their velocities, independent of possible interactions. Secondly, collisions between proximate particles are resolved using a stochastic approach. Macroscopic flow properties are obtained from particle microscopic properties averaged over many time steps. To accurately simulate multi-species flows, the time step should be smaller than the mean collision time of the particles, the cell size should be smaller than the mean free path, and a sufficient number of simulated particles, typically 10 to 20 per cell for each species, should be used to ensure a precise statistical description of the flow.

The computational code used is SPARTA~\cite{plimpton_direct_2019} (version of January 20, 2025), an open source code created by Sandia National Laboratories under the GNU Public License. SPARTA features collision calculations using Bird's~\cite{bird_molecular_1976} no time counter (NTC) algorithm and adaptive orthogonal grids with efficient parallelization and load balancing. The variable soft sphere (VSS) model is used as the collision model. Since electrons have an extremely small mass, approximately one thousandth of the mass of a proton, they have high velocities compared to other particles. Consequently, electron movement calculations would require an extremely small time step. Therefore, the ambipolar diffusion model~\cite{bird_nonequilibrium_1987, ozawa_development_2008, kaplan_analysis_2022} is used, assuming electrons move with ions while maintaining charge quasi-neutrality. That is, electron velocity information is stored with the corresponding ion particles, and separate simulated particles for the ambipolar diffusion electrons are used only during collision calculations. The total collision energy (TCE) model determines reaction occurrence by incorporating the modified Arrhenius equation at the molecular level. Energy distribution during collisions is modeled using the Borgnakke-Larsen model~\cite{borgnakke_statistical_1975}, considering translational, rotational, and vibrational energy modes.

In the present simulation case involving weakly ionized plasma, trace species such as electrons have mole fractions on the order of $10^{-4}$ or lower. Meeting the above numerical conditions, particularly the particles per cell requirement, while obtaining electron number density fields is challenging, because the number of simulated particles for each species is proportional to their mole fractions in the conventional DSMC framework. Satisfying this requirement for trace species, such as electrons, therefore necessitates a proportionally large number of simulated particles for the major species, making the computational load enormous. Therefore, in this paper, a reactive SWS is developed to obtain the trace species. Since conventional DSMC can give macroscopic flow fields but not trace species, it was used to validate the bulk flow fields obtained with the SWS.
  
\subsection{Conservative Species Weighting Scheme}
\label{sec:sws}

The development of this method required implementing a new collision scheme in the SPARTA code. In this method, a weighting factor $w_i$ is assigned to each chemical species $i$, so that all simulated particles of species $i$ have the same weight. This weighting factor takes values such that $0 < w_i \leq 1$ and modifies the number of real molecules represented by simulated particles. The final number of real molecules is given by the product of the global weighting factor $F_N$, which is the ratio of real molecules to simulated particles, and the species weight $w_i$.

Conventional DSMC corresponds to the case where $w_i = 1.0$ for all chemical species. Setting a low $w_i$ results in a smaller product $w_i F_N$, meaning fewer real molecules are contained in each simulated particle of species $i$. This implies that more simulated particles are used in DSMC calculations to obtain the same number of real molecules. The choice of $w_i$ therefore affects only the number of simulated particles, not the physical number density of the species. Therefore, setting low weights for trace species can increase their number of simulated particles. The species weighting scheme is thus introduced for systems with large number density differences between chemical species. While this method resolves the disparity in simulated particle numbers between species, it requires additional considerations for pair selection, momentum, and energy conservation beyond conventional DSMC. Furthermore, special attention to mass conservation is needed due to chemical reactions.

\subsection{Collision count, collision pair selection, and collision probability}
\label{sec:collision_process}

This section first explains the modifications related to particle collisions. In the conventional NTC model, the number of attempted collision pairs $N_\mathrm{c}$ is determined by the number of molecules represented by one particle $F_N$, the number of simulated particles $N$, the time step $\Delta t$, the collision cross section $\sigma_\mathrm{t}$, the relative velocity of colliding particles $g_\mathrm{r}$, and the cell volume $V_\mathrm{cell}$:
\begin{equation}
\label{eq:ntc_num_collide}
N_\mathrm{c} = \frac{1}{2}N(N-1)F_N(\sigma_\mathrm{t} g_\mathrm{r})_\mathrm{max}\Delta t/V_\mathrm{cell}.
\end{equation}

With species weighting introduced, the number of real molecules in a cell is expressed as the sum of weighting factors $F_N\sum_j w_{i,j}$ of each particle $j$. Therefore, the modified number of collision attempts can be expressed as:
\begin{equation}
\label{eq:ntc_num_collide_1st_mod}
N_\mathrm{c} = \frac{1}{2}(N-1)F_N(\sigma_\mathrm{t} g_\mathrm{r})_\mathrm{max}\Delta t\sum_j w_{i,j}/V_\mathrm{cell}.
\end{equation}

After the NTC algorithm calculates the number of collision pairs, candidate particles are randomly selected from the particles in the cell. Each formed pair is verified by the acceptance-rejection method according to collision probability $P_\mathrm{c}$, which depends on the particles' collision cross section and relative velocity. This probability is expressed as:
\begin{equation}
  \label{eq:DSMC_collision_probability}
  P_\mathrm{c} = \frac{\sigma_\mathrm{t} g_\mathrm{r}}{(\sigma_\mathrm{t} g_\mathrm{r})_\mathrm{max}},
\end{equation}
where ${(\sigma_\mathrm{t} g_\mathrm{r})_\mathrm{max}}$ is the maximum value of $\sigma_\mathrm{t} g_\mathrm{r}$, which is updated during the collision process in each cell. It is typically initialized to zero at the beginning of each time step. This parameter is reset to zero every 1000 collision attempts in the present simulation. Although the number of collision attempts in the NTC algorithm does not correspond to the actual collision numbers, particle pairs with higher $\sigma_\mathrm{t} g_\mathrm{r}$ are selected for collision with higher probability, ensuring accurate flow behavior representation.

When using the species weighting scheme, the number of simulated particles is no longer proportional to the mixture composition. Therefore, since pair selection is a random process, more trace species may be selected for collision compared to conventional DSMC, introducing a bias in the flow behavior. Since these trace species collisions involve only a small number of real molecules due to their low weights, this bias does not influence the simulation accuracy significantly, especially when an appropriate balance between trace and major simulated particles is maintained, with a ratio of at least 10 major particles per trace particle. For the same reason, Boyd~\cite{boyd_conservative_1996} recommends using the species weighting scheme only for species with mole fractions below 0.1 and selecting the ratio of weights between trace and major species to be approximately equal to their mole fraction ratio.

To address these biases and accurately simulate flow collisions with differently weighted particles, pair selection methods have been developed, such as the method described by Charton et al.~\cite{charton_species_2025}, and those used in the stochastic weighted particle method by Hong and Gallis~\cite{hong_improved_2024} and Massaro et al.~\cite{massaro_efficient_2024}. This study adopts the method referred to as the weight maximum filter, as used by Charton et al.~\cite{charton_species_2025}.

Here, we briefly outline the required changes from conventional DSMC and provide an explanation of each.

Firstly, to correct the collision probability between two particles with different weights, considering the collision between particle $a$ representing chemical species $\mathrm{A}$ with weight $w_\mathrm{A}$ and particle $b$ representing chemical species $\mathrm{B}$ with weight $w_\mathrm{B}$, Eq.~\eqref{eq:DSMC_collision_probability} is modified as follows:
\begin{equation}
  \label{eq:DSMC_collision_probability_maxfilter}
  P_\mathrm{c} = \frac{\sigma_\mathrm{t} g_\mathrm{r}}{(\sigma_\mathrm{t} g_\mathrm{r})_\mathrm{max}}\frac{\max{(w_\mathrm{A},w_\mathrm{B})}}{\max(w_i)},
\end{equation}
where $\mathrm{max}(w_i)$ is the maximum value of the weighting factor $w_i$ across all the chemical species in the cell. Consequently, the probability of accepting collisions of particles with larger weights increases, while the collision probability of particles with smaller weights decreases. Secondly, to conserve the total number of physical particle collisions when using Eq.~\eqref{eq:DSMC_collision_probability_maxfilter}, the total number of collision attempts must be increased. Compared to Eq.~\eqref{eq:ntc_num_collide_1st_mod}, Eq.~\eqref{eq:ntc_num_collide} is modified as follows, where $N \max(w_i) \geq \sum_j w_{i,j}$:
\begin{equation}
  \label{eq:max_filter_num_collide}
  N_\mathrm{c} = \frac{1}{2}N (N-1)F_N \max(w_i) (\sigma_\mathrm{t} g_\mathrm{r})_\mathrm{max}\Delta t/V_\mathrm{cell}.
\end{equation}
This approach is adopted throughout the present SWS simulations.

\subsection{Momentum conservation}
\label{sec:momentum_conservation}

\begin{figure*}[hbtp]
  \centering
  \includegraphics[width=0.9\textwidth]{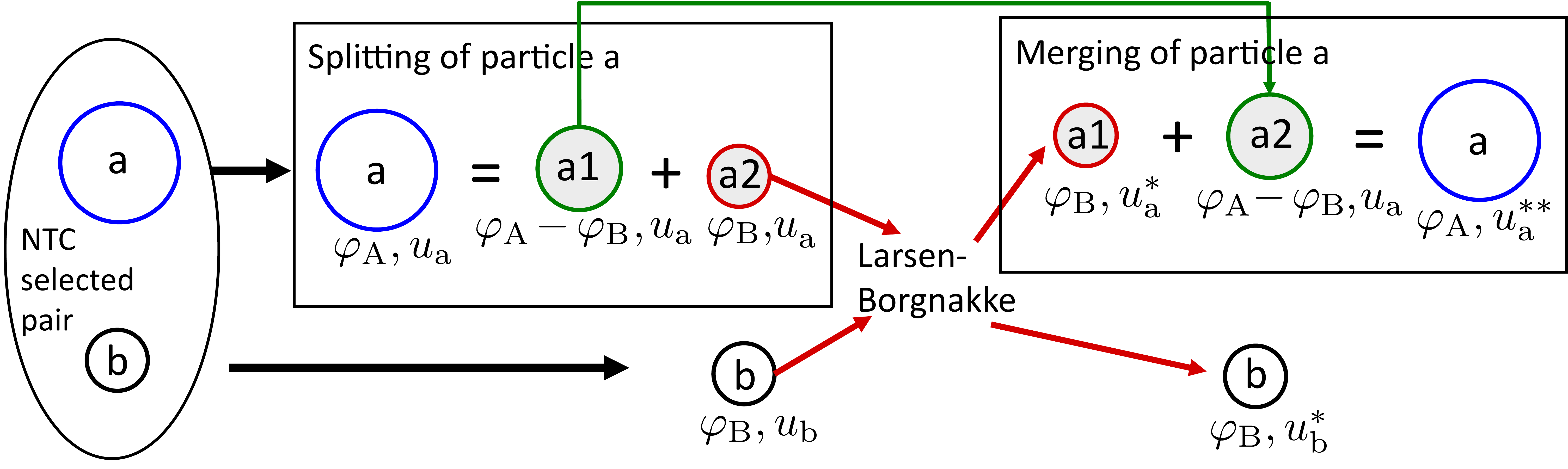}
  \caption{Schematic of the split-merge model applied to the elastic collision of particles $a$ and $b$ having velocities $u_a$ and $u_b$ respectively, and representing $\varphi_\mathrm{A}$ and $\varphi_\mathrm{B}$ number of physical particles with $w_\mathrm{A} > w_\mathrm{B}$. The post-collision properties are denoted with the symbol ${}^{*}$, and the merged particle with the symbol ${}^{**}$.}
  \label{fig:split-merge_schematic}
\end{figure*}

In this section, the momentum and energy conservation using Boyd's~\cite{boyd_conservative_1996} split-merge method is explained through a one-dimensional collision. Figure~\ref{fig:split-merge_schematic} shows a schematic of a collision between particles with different weights. Consider the collision between particle $a$ representing species A with weight $w_\mathrm{A}$ and particle $b$ representing species $\mathrm{B}$ with weight $w_\mathrm{B}$. The masses of molecules represented by particles $a$ and $b$ are $m_\mathrm{A}$ and $m_\mathrm{B}$,
 respectively, and their pre-collision velocities are $u_a$ and $u_b$. When $w_\mathrm{A} > w_\mathrm{B}$, i.e., when B is a trace species, the weight ratio is set as $\phi = w_\mathrm{B}/w_\mathrm{A}$. The collision momentum is expressed as:
\begin{equation}
  \label{eq:SWS_mom_1}
p_\mathrm{m} = F_N w_\mathrm{A} m_\mathrm{A} u_a + F_N w_\mathrm{B} m_\mathrm{B} u_b.
\end{equation}

In the split approach, from particle $a$ containing $\varphi_\mathrm{A} = F_N w_\mathrm{A}$ real molecules, only $\varphi_\mathrm{B} = F_N w_\mathrm{B}$ real molecules are extracted to actually collide and exchange energy with particle $b$. The remaining $\varphi_\mathrm{A} - \varphi_\mathrm{B} = F_N(w_\mathrm{A} - w_\mathrm{B})$ real molecules do not collide. Considering this, Eq.~\eqref{eq:SWS_mom_1} is modified as:
\begin{equation}
p_\mathrm{m} = F_N w_\mathrm{A}({(1 - \phi)m_\mathrm{A} u_a + \phi(m_\mathrm{A} u_a + m_\mathrm{B} u_b)}).
\end{equation}

The collision of $F_N w_\mathrm{A}\phi = F_N w_\mathrm{B}$ real molecules is processed using the Borgnakke-Larsen method as in conventional DSMC, yielding post-collision velocities $u_a^*$ and $u_b^*$. Here, an asterisk~$*$ indicates the state obtained after the Borgnakke-Larsen method, and two asterisks~$**$ indicate the state after modification using the split-merge method. Substituting the post-collision velocities into the momentum expression gives:
\begin{equation}
  \label{eq:SWS_mom_coll_pre_SM}
p_\mathrm{m}  = F_N w_\mathrm{A}({(1 - \phi)m_\mathrm{A} u_a + \phi(m_\mathrm{A} u_a^* + m_\mathrm{B} u_b^*)}).
\end{equation}

The final step is to merge the split particle $a$ to obtain the final post-collision velocity $u_a^{**}$. To conserve momentum during the collision process, Eq.~\eqref{eq:SWS_mom_coll_pre_SM} should equal:
\begin{equation}
p_\mathrm{m} = F_N w_\mathrm{A} m_\mathrm{A} u_a^{**} + F_N w_\mathrm{B} m_\mathrm{B} u_b^*.
\end{equation}
This yields $u_a^{**}$ explicitly satisfying momentum conservation:
\begin{equation}
u_a^{**} = (1-\phi)u_a + \phi u_a^*.
\end{equation}

\subsection{Energy conservation}
\label{sec:energy_conservation}

However, while this method conserves momentum, it does not conserve translational energy. The pre-collision $E_\mathrm{pre}$ and post-collision $E_\mathrm{post}$ translational energies are given by:
\begin{equation}
E_\mathrm{pre} = F_N w_\mathrm{A} \left(\frac{1}{2}m_\mathrm{A} u_a^2\right) + F_N w_\mathrm{B} \left(\frac{1}{2}m_\mathrm{B} u_b^2 \right),
\end{equation}
\begin{equation}
E_\mathrm{post} = F_N w_\mathrm{A} \left(\frac{1}{2}m_\mathrm{A} u_a^{**2}\right) + F_N w_\mathrm{B} \left(\frac{1}{2}m_\mathrm{B}u_b^{*2} \right).
\end{equation}
Therefore, the loss amount $\Delta E_{\mathrm{loss}}$ is:
\begin{equation}
  \label{eq:split_merge_E_loss}
  \Delta E_{\mathrm{loss}} = F_N w_\mathrm{A} \left( \frac{1}{2}m_\mathrm{A}\phi(1-\phi)(u_a - u_a^*)^2 \right).
\end{equation}
When $w_\mathrm{A} = w_\mathrm{B}$, $\phi = 1.0$ and $\Delta E_{\mathrm{loss}} = 0$, resulting in perfect energy conservation. However, with differently weighted particles, $\Delta E_{\mathrm{loss}} > 0$, causing continuous energy loss from the system. 
This issue is addressed by tracking energy loss in each cell: when energy is lost in collisions according to Eq.~\eqref{eq:split_merge_E_loss}, it is recorded and added to the translational energy of subsequent collisions. While this approach conserves energy globally across the system, energy is not strictly conserved in each individual collision. Energy addition occurs in collisions between particles with maximum weights, namely, collisions between major species particles (collisions not involving trace species simulated particles), because distributing the tracked energy loss to trace species may give physically unreasonable velocities.

\subsection{Internal energy conservation}
\label{sec:internal_energy_conservation}

In systems involving hypersonic flows of high temperature air, the internal energy should be accounted for in the simulation. The present DSMC calculations treat the vibrational mode discretely and the rotational mode continuously. For these internal energy modes as well, collisions between simulated particles with different weights are handled using the method detailed in Sec.~\ref{sec:energy_conservation}.

Let us consider a collision between particle $a$ (species A) and particle $b$ (species B) again. In addition to the variables defined in Sec.~\ref{sec:energy_conservation}, assume that particles $a$ and $b$ possess energies $e_{\mathrm{int},a}$ and $e_{\mathrm{int},b}$, respectively, in a given internal energy mode. When $w_\mathrm{A} = w_\mathrm{B}$, internal energy exchange can be directly computed using the Borgnakke-Larsen method without any issues. Now, let us consider the case of unequal weights, specifically $w_\mathrm{A} > w_\mathrm{B}$. For continuous internal energy modes, the split-merge method can be applied in the same manner as for translational energy:

\begin{equation}
\label{eq:eint_split-merge}
e_{\mathrm{int},a}^{**} = (1-\phi)e_{\mathrm{int},a} + \phi e_{\mathrm{int},a}^*.
\end{equation}

However, this split-merge approach is not directly applicable to discrete energy modes, as it would result in non-discrete energy states. For instance, if the split-merge procedure were applied to the vibrational energy of particle $a$ in the above example, its post-collision energy $e_{\mathrm{int},a}^{**}$ would be modified to a value that no longer aligns with the allowable discrete levels. Therefore, the split-merge internal energy exchange is rejected for the discrete modes. Rejecting the exchange in these specific instances has a negligible effect on the overall relaxation rates. This approach is justified because the weight ratio $\phi = w_\mathrm{B}/w_\mathrm{A}$ is generally very small, and such energy exchange collisions involving differently weighted particles are rare. When $\phi \ll 1$, Eq.~\eqref{eq:eint_split-merge} implies that $e_{\mathrm{int},a}^{**} \approx e_{\mathrm{int},a}$, which poses no significant concern from an energy conservation point of view.

 The conservation procedures from Sec.~\ref{sec:momentum_conservation} to Sec.~\ref{sec:internal_energy_conservation} were implemented in SPARTA and validated for non-reactive flows by Charton et al.~\cite{charton_species_2025}.

\subsection{Chemical reactions}
\label{sec:chemical_reactions}

In conventional DSMC, when two particles collide and a reaction is determined to occur by the TCE algorithm, two (exchange reaction, associative ionization reaction, etc.) or three (dissociation reaction, etc.) simulated particles of the product species are produced depending on the reaction type. Since reactant particles and product particles represent the same number of real molecules, mass is naturally conserved. However, when using the species-unique weight, special consideration is necessary because the weights $w_i$ of reactant and product particles differ.

\begin{figure*}[hbtp]
  \centering
  \includegraphics[width=0.9\textwidth]{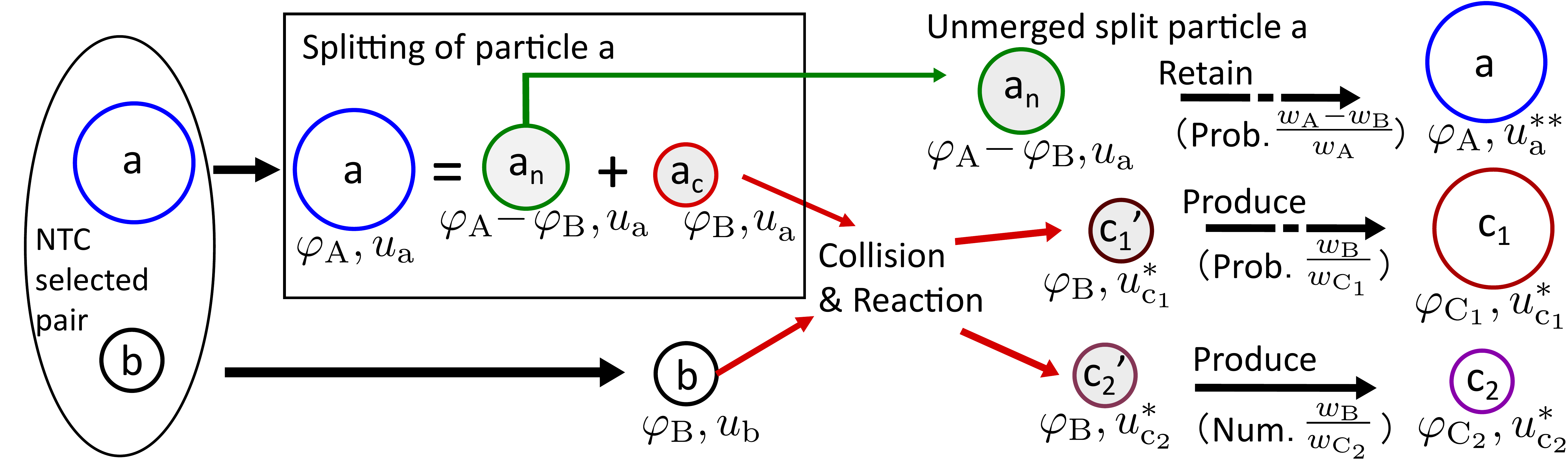}
  \caption{Schematic of the split-merge model for a reactive collision between particles $a$ and $b$ (with velocities $u_a$ and $u_b$), representing $\varphi_\mathrm{A}$ and $\varphi_\mathrm{B}$ physical molecules $(w_\mathrm{A} > w_\mathrm{B}, w_\mathrm{r,min} = w_\mathrm{B})$, and producing two product species, $\mathrm{C}_1$ and $\mathrm{C}_2$. In this schematic, we assume $w_{\mathrm{C}_1} \geq w_\mathrm{B}$ and $w_{\mathrm{C}_2} \leq w_\mathrm{B}$.}
  \label{fig:chemical_split-merge_schematic}
\end{figure*}

Based on the split-merge method, only the amount corresponding to the particle with the minimum weight reacts, and an equal amount of products is produced. Therefore, the final number of simulated particles produced for all reactions involving minor species depends on the minimum weight $w_\mathrm{r,min}$ involved in the collision process. The occurrence of a reaction is determined by the TCE routine, as in conventional DSMC.
Figure~\ref{fig:chemical_split-merge_schematic} illustrates the split-merge treatment for reactive collisions between particles with different weights. The detailed chemical reaction process is described below.

\begin{figure*}[htbp]
  \centering
  \begin{subfigure}[t]{0.48\textwidth}
    \centering
    \includegraphics[width=\textwidth]{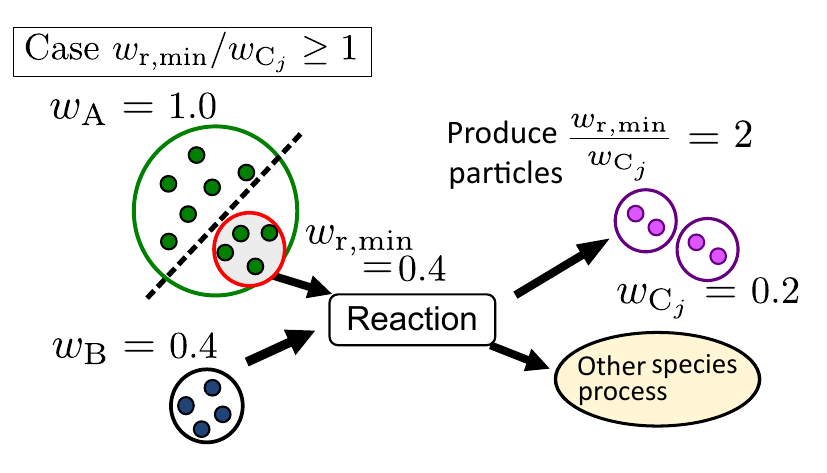}
    \caption{Case $w_\mathrm{r,min}/w_{\mathrm{C}_j} \geq 1$}
    \label{fig:schematic_product_1}
  \end{subfigure}
  \hfill
  \begin{subfigure}[t]{0.48\textwidth}
    \centering
    \includegraphics[width=\textwidth]{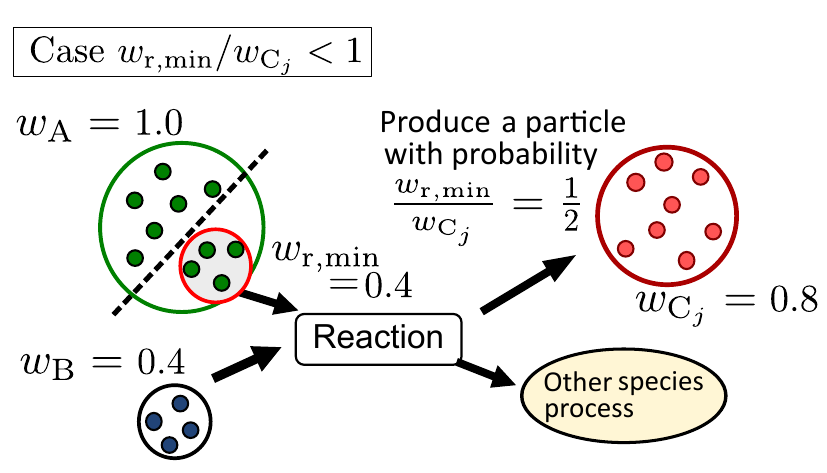}
    \caption{Case $w_\mathrm{r,min}/w_{\mathrm{C}_j} < 1$}
    \label{fig:schematic_product_2}
  \end{subfigure}
  \caption{Schematic illustrating the reactive collision of split particles and the resulting product species production process: (a) production of an integer number of particles when $w_\mathrm{r,min}/w_{\mathrm{C}_j} \geq 1$, and (b) probabilistic production of a single particle when $w_\mathrm{r,min}/w_{\mathrm{C}_j} < 1$.}
  \label{fig:schematic_product_cases}
\end{figure*}

As we discussed in Sec.~\ref{sec:momentum_conservation}, let us consider the collision reaction between particle $a$ and trace species simulated particle $b$, producing particle $c_j$ representing $j$-th product species $\mathrm{C}_j$. The minimum weighting factor among the reactant particles is defined as $w_\mathrm{r,min} = \min(w_\mathrm{A}, w_\mathrm{B})$. In this example, we let $w_\mathrm{r,min} = w_\mathrm{B}$, and then $F_N w_\mathrm{r,min} = F_N w_\mathrm{B}$ molecules participate in the reaction. The production process for product-species simulated particles changes based on the ratio between $w_\mathrm{r,min}$ and $w_{\mathrm{C}_j}$:

\begin{enumerate}
\item Case $w_\mathrm{r,min}/w_{\mathrm{C}_j} \geq 1$:\\
The number of real molecules represented by particle $c_j$ is less than the number of real molecules involved in the reaction, so particle $c_j$ is replicated into $w_\mathrm{r,min}/w_{\mathrm{C}_j}$ particles. An example of this case, shown in Fig.~\ref{fig:schematic_product_1}, is when the reactant species weights are $w_\mathrm{A}=1.0$ and $w_\mathrm{B}=0.4$ (giving $w_\mathrm{r,min}=0.4$), and the product species weight is $w_{\mathrm{C}_j}=0.2$, resulting in a ratio of $2$. In this case, two product particles are certainly produced to maintain mass conservation. 
One idea for assigning velocities to these particles is to give all of them the exact same velocity vector as clones. Galitzine and Boyd~\cite{galitzine_adaptive_2015} used a similar process of producing cloned particles during computational optimization. However, if the particles have the same velocity vector, they do not collide with each other. Galitzine and Boyd note that this issue is not important for relatively high gas density, where the number of collisions with other particles is high enough.
To avoid this problem, in the present implementation, each product particle independently undergoes VSS velocity scattering using the same pre-collision relative velocity but with an independently drawn random scattering angle, yielding distinct post-collision velocity vectors for each particle. As a drawback of this independent scattering approach, the other computed velocity update is discarded after each scattering, so that energy conservation holds statistically but not in each individual collision.

\item Case $w_\mathrm{r,min}/w_{\mathrm{C}_j} < 1$:\\
The number of real molecules represented by particle $c_j$ is greater than the number of real molecules involved in the reaction. Therefore, generating a single particle $c_j$ would result in over-representation of species $\mathrm{C}_j$ molecules. To address this via a probabilistic mass conservation approach, particle $c_j$ is produced with probability $w_\mathrm{r,min}/w_{\mathrm{C}_j}$ and not produced otherwise. This scenario is illustrated in Fig.~\ref{fig:schematic_product_2}, where $w_\mathrm{r,min}=0.4$ and a larger product species weight of $w_{\mathrm{C}_j}=0.8$ are given, resulting in a ratio of $1/2$. Consequently, a single product particle is produced with a $1/2$ probability to ensure macroscopic mass conservation.

\end{enumerate}

The handling of reactant particles also follows this probabilistic mass conservation approach. In a reaction between particles with different weights, the physical number of reacting molecules is governed by the minimum weight $w_\mathrm{r,min}$. Consequently, if a reactant particle $a$ or $b$ has a weight $w_\mathrm{r,max} = \max(w_\mathrm{A}, w_\mathrm{B}) > w_\mathrm{r,min}$, it represents more physical molecules than those consumed in the single reaction event. The number of physical molecules that should remain corresponds to $F_N(w_\mathrm{r,max} - w_\mathrm{r,min})$. To maintain mass conservation, this reactant particle is not unconditionally maintained or deleted. Instead, it is retained in the simulation with a survival probability of $(w_\mathrm{r,max} - w_\mathrm{r,min})/w_\mathrm{r,max}$, which represents the fraction of molecules that did not interact. Otherwise, the particle is deleted. The reactant particle possessing the minimum weight $w_\mathrm{r,min}$ is always deleted as it is fully consumed.

Regarding this probabilistic mass conservation approach, as mentioned in Section~\ref{sec:intro}, an alternative would be to generate new simulated particles with weights corresponding to the produced or remaining amounts. However, this approach can lead to a proliferation of small-weight particles that increase computational cost without significant improvement in results. Other schemes, such as the stochastic weighted particle method, address this by adding a reduction step outside the collision routine; however, such procedures are non-trivial and may introduce additional complexity and discrepancy. In the present study, we therefore adopt the probabilistic mass conservation approach. Although mass is conserved on average over a large number of reactive collisions rather than for each individual collision, this limitation does not lead to a significant discrepancy in the current application. Through this handling, the system's mass, momentum, and energy are conserved on average.

\subsection{Verification of the chemical reaction treatment}
\label{subsec:verification_reaction}
 The present method was additionally implemented in SPARTA and validated using two single-cell simulation cases with chemical reactions.

The first case considered the dissociation reaction $\mathrm{N_2 + N_2 \to N + N + N_2}$ in a two-dimensional single cell of 0.01 $\times$ 0.01 m with unit depth, at $T = 20000$ K, with an initial mole fraction of $X_{\mathrm{N_2}} = 1.0$, number density $1.0 \times 10^{23}$ m$^{-3}$, and the global weighting factor $F_N = 1.0 \times 10^{16}$. Simulated particles of $\mathrm{N_2}$ were assigned species weight $w_{\mathrm{N_2}} = 1$, while the product $\mathrm{N}$ atom was assigned a ten times smaller species weight $w_{\mathrm{N}} = 0.1$, representative of the trace species treatment used in practical hypersonic flow computations. The second case examined the exchange reaction $\mathrm{N_2 + O \to NO + N}$ under the same geometry, thermal conditions, number density and the global weighting factor $F_N$, with initial mole fractions $X_{\mathrm{N_2}} = 0.9$, $X_{\mathrm{O}} = 0.1$, product species weights $w_{\mathrm{NO}} = 1, w_{\mathrm{N}} = 0.1$, and reactant species weights $w_{\mathrm{N_2}} = 1, w_{\mathrm{O}} = 0.1$. The reaction rate parameters follow the set listed in Table~\ref{tab:rate_parameters_mole}. In both cases, the energy change due to reactions was set to zero to allow a clear validation of the SWS energy conservation without inputting energy changes due to the reactions themselves.

\begin{figure*}[htbp]
  \centering
  \begin{subfigure}[t]{0.48\textwidth}
    \centering
    \includegraphics[width=\textwidth]{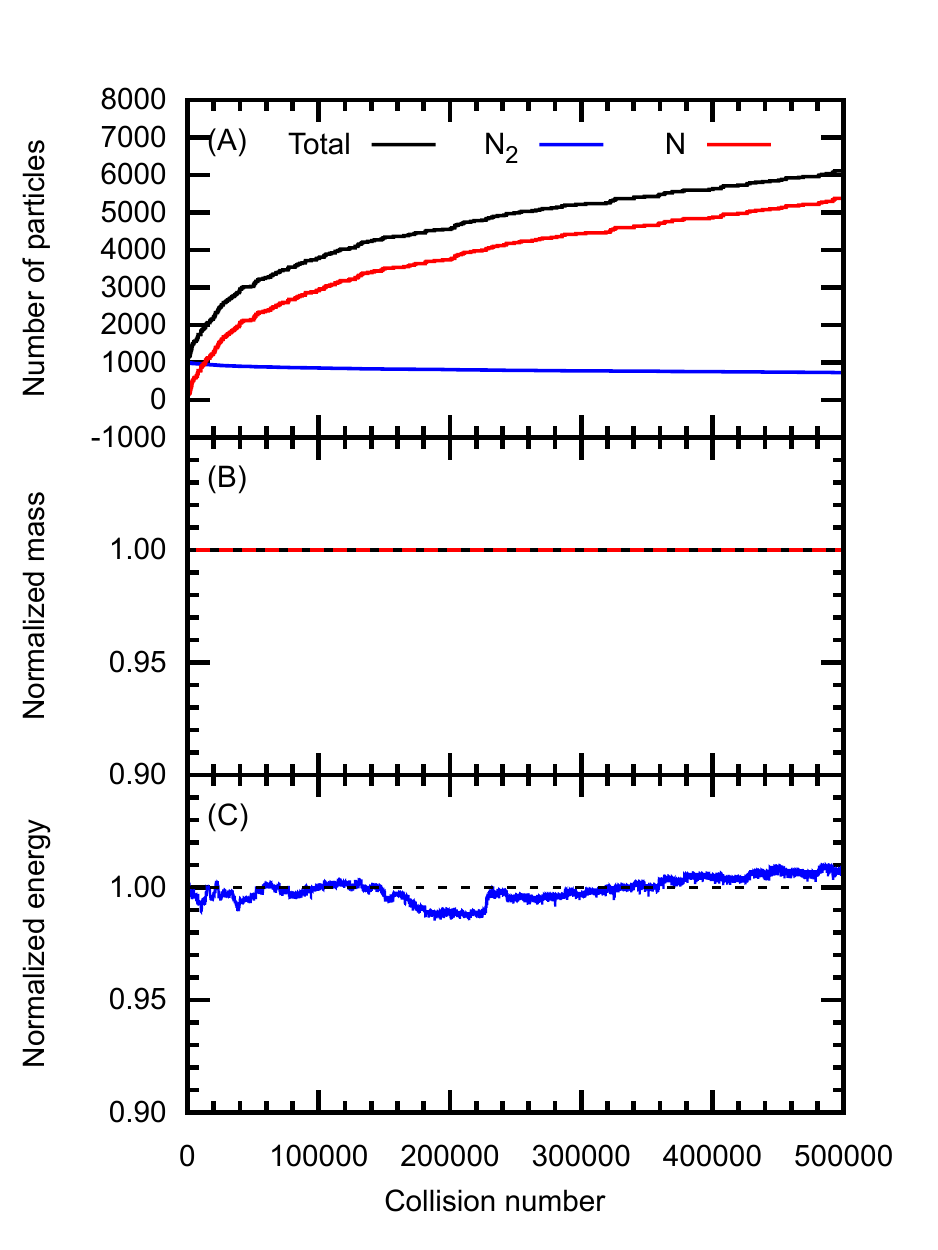}
    \caption{Dissociation reaction $\mathrm{N_2 + N_2 \to N + N + N_2}$ ($w_{\mathrm{N_2}}=1$, $w_{\mathrm{N}}=0.1$).}
    \label{conservation_1}
  \end{subfigure}
  \hfill
  \begin{subfigure}[t]{0.48\textwidth}
    \centering
    \includegraphics[width=\textwidth]{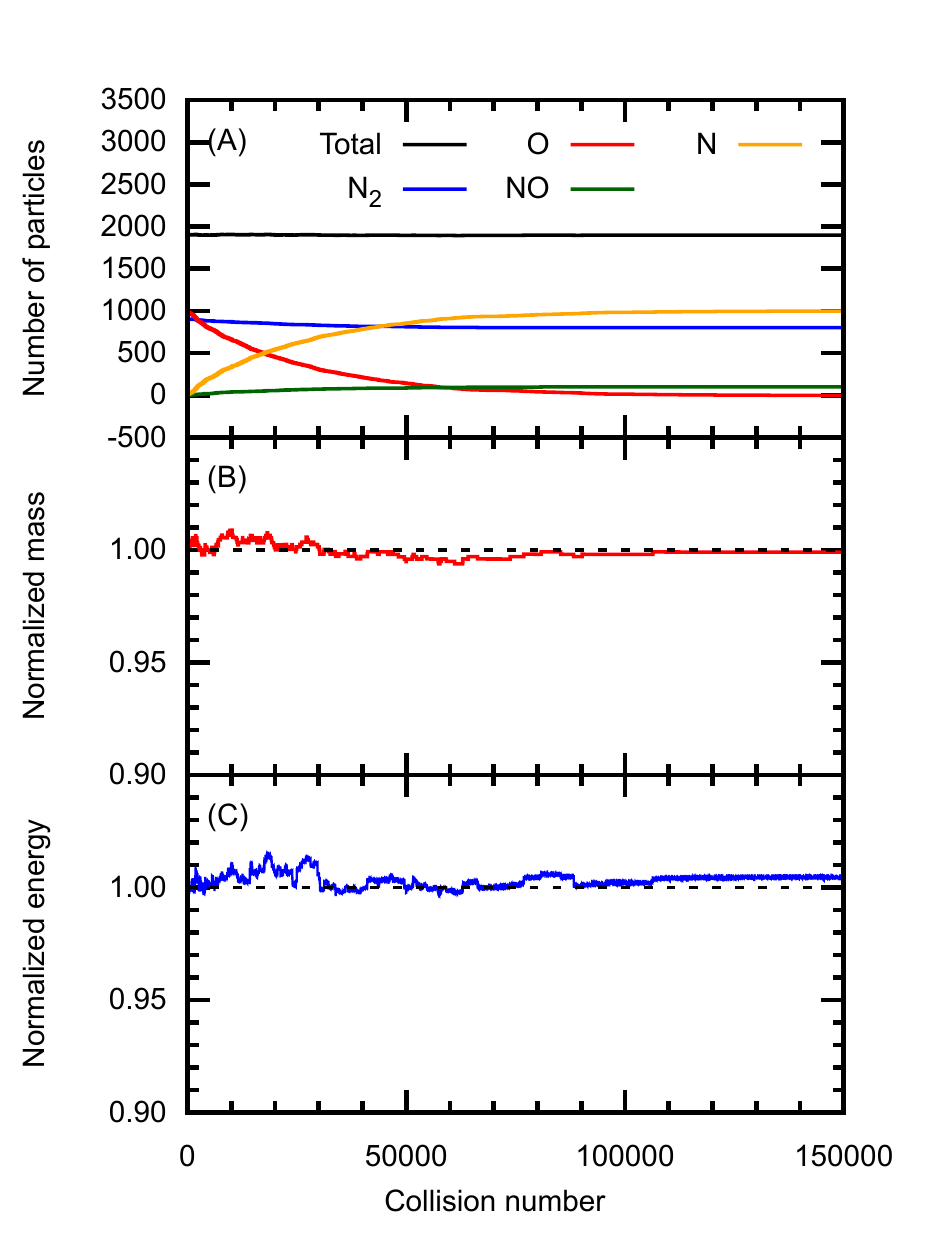}
    \caption{Exchange reaction $\mathrm{N_2 + O \to NO + N}$ ($w_{\mathrm{N_2}} = w_{\mathrm{NO}}=1$, $w_{\mathrm{N}}=w_{\mathrm{O}}=0.1$).}
    \label{conservation_2}
  \end{subfigure}
  \caption{Single-cell simulations for chemical-reaction cases with species-unique weights. Each figure shows the evolution within a single time step. (A) Number of simulated particles per species. (B) Total mass normalized by its initial value. (C) Total translational energy of the simulated particles normalized by its initial value.}
  \label{fig:conservation_reaction}
\end{figure*}

The results are shown in Fig.~\ref{fig:conservation_reaction}. Each figure contains three graphs with respect to collision count in a single time step. Graph (A) shows the evolution of the number of simulated particles for each species, confirming that species conversion proceeds as reactions occur. Graph (B) shows the total system mass normalized by its initial value. Graph (C) shows the normalized total translational energy carried by simulated particles.
In graph (A), for the dissociation case shown in Fig.~\ref{conservation_1}, the number of $\mathrm{N_2}$ particles decreases monotonically, whereas the number of $\mathrm{N}$ particles increases. The integer weight ratio $w_{\mathrm{N_2}}/w_{\mathrm{N}} = 10$ means that each of the two product $\mathrm{N}$ atoms is represented by 10 simulated particles, so that 20 simulated particles of $\mathrm{N}$ are produced per reaction event. For the exchange case shown in Fig.~\ref{conservation_2}, the numbers of $\mathrm{O}$ and $\mathrm{N}$ particles change at ten times the rate of those of $\mathrm{N_2}$ and $\mathrm{NO}$, which corresponds to the weight ratio $w_{\mathrm{N_2}}/w_{\mathrm{O}} = w_{\mathrm{NO}}/w_{\mathrm{N}} = 10$.
In graph (B), the normalized mass remained exactly at unity for the dissociation case of Fig.~\ref{conservation_1}. Since the weight ratios $w_\mathrm{r,min}/w_{\mathrm{C}_j}$ are integers, no probabilistic process is involved in particle production, and each reaction event conserves mass exactly. For the exchange reaction case of Fig.~\ref{conservation_2}, the reactant $\mathrm{O}$ has weight $w_{\mathrm{O}} = 0.1$ while the collision partner $\mathrm{N_2}$ and the product $\mathrm{NO}$ have the weight $w_{\mathrm{N_2}} = w_{\mathrm{NO}} = 1$; consequently, the ratio $w_\mathrm{r,min}/w_{\mathrm{C}_j}$ is non-integer, and a probabilistic process arises for both reactant and product particles.
In graph (C), the total translational energy of the simulated particles fluctuates around unity for both cases. The fluctuation is due to the probabilistic process of the reaction events.
In the dissociation case shown in Fig.~\ref{conservation_1}, even though the mass is exactly conserved, the energy is not perfectly conserved due to the velocity perturbation applied to cloned particles to prevent identical velocities, as described in Sec.~\ref{sec:chemical_reactions}. 
In the exchange reaction case shown in Fig.~\ref{conservation_2}, the energy fluctuation is linked to the probabilistic creation and deletion of reactant and product particles: as particles are added or deleted, their associated kinetic energy enters or leaves the system. Consequently, the energy fluctuations follow a pattern similar to that of the total mass variations for reactants and products. 
In both graphs (B) and (C), fluctuations of approximately $\pm 2\%$ were observed. A mole fraction of 0.1 was used for these verification cases, which is high for typical trace species and was chosen so
that any deviation due to the species weighting would be visible. 

Despite this mole fraction, the observed fluctuations are limited, and these results validate the present implementation of the species-unique weight collision and reaction model in the code.

\begin{figure*}[htbp]
  \centering
  \begin{subfigure}[t]{0.48\textwidth}
    \centering
    \includegraphics[width=\textwidth]{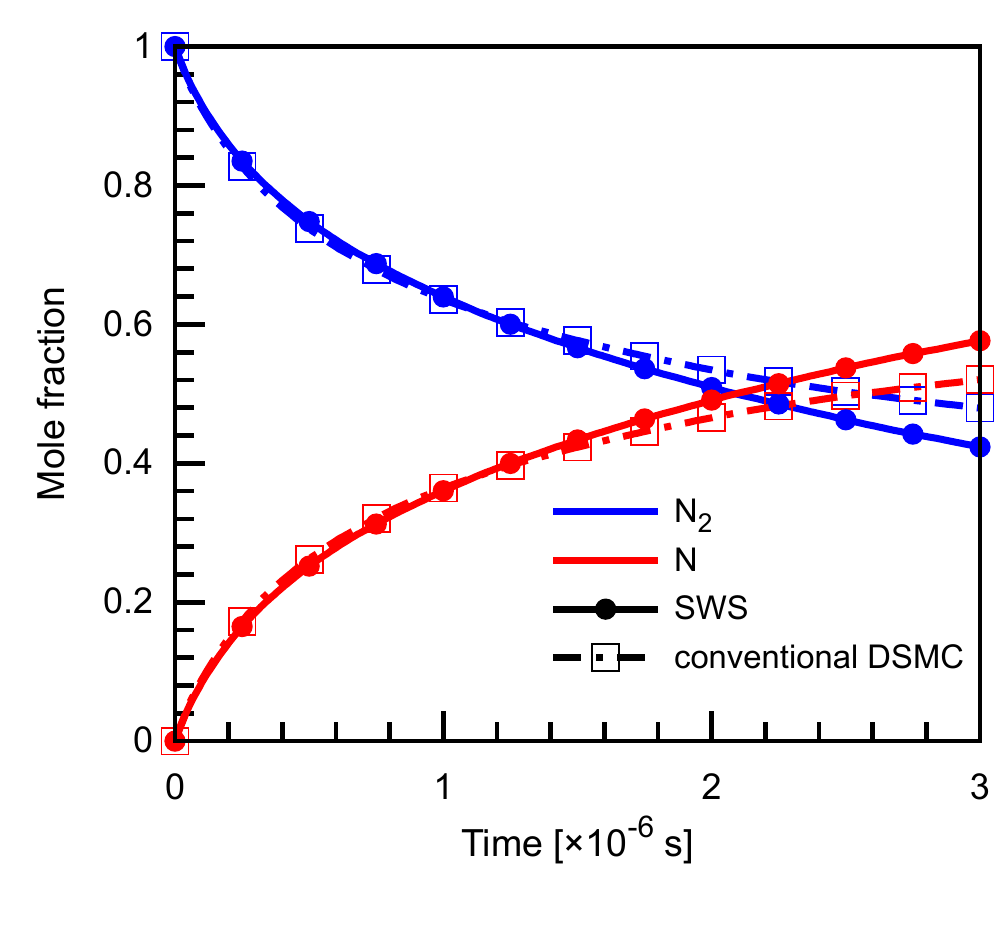}
    \caption{Dissociation reaction $\mathrm{N_2 + N_2 \to N + N + N_2}$ ($w_{\mathrm{N_2}}=1$, $w_{\mathrm{N}}=0.1$).}
    \label{mole_frac_rate_1}
  \end{subfigure}
  \hfill
  \begin{subfigure}[t]{0.48\textwidth}
    \centering
    \includegraphics[width=\textwidth]{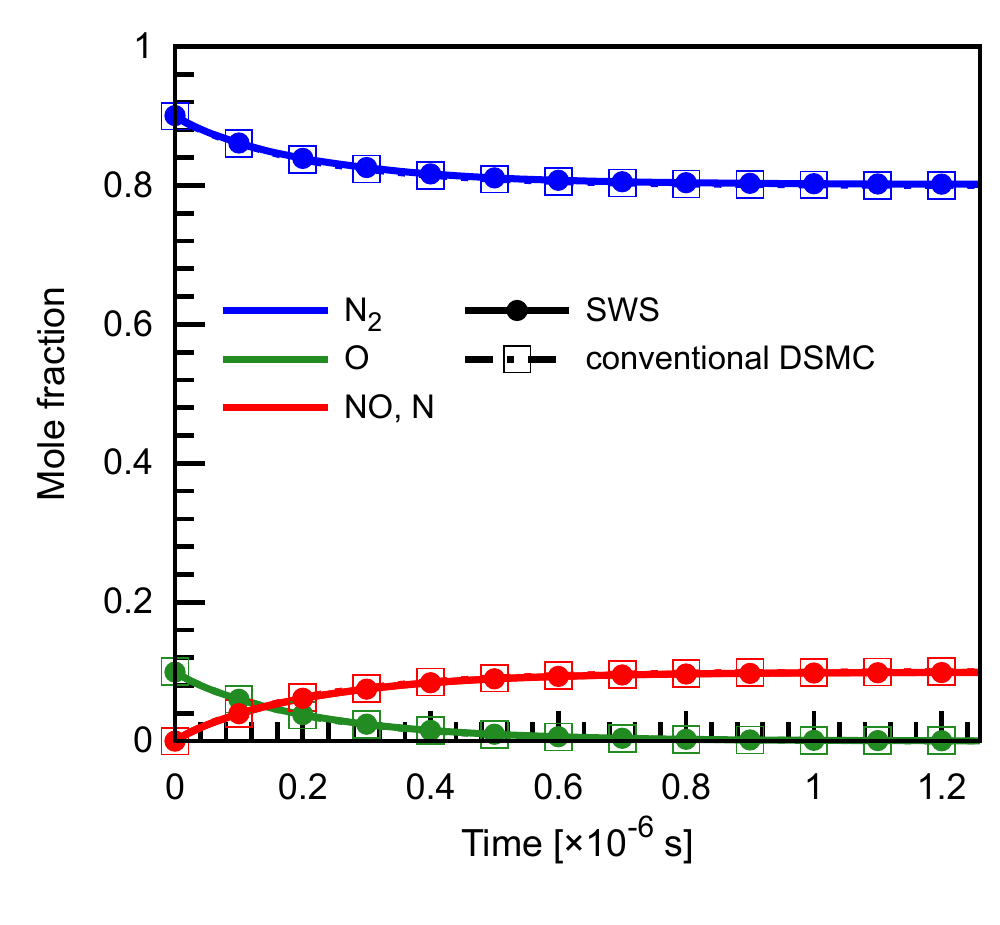}
    \caption{Exchange reaction $\mathrm{N_2 + O \to NO + N}$ ($w_{\mathrm{N_2}} = w_{\mathrm{NO}}=1$, $w_{\mathrm{N}}=w_{\mathrm{O}}=0.1$).}
    \label{mole_frac_rate_2}
  \end{subfigure}
  \caption{Single-cell simulations for chemical-reaction cases comparing the SWS and conventional DSMC methods over longer times: (a) dissociation reaction and (b) exchange reaction. Each figure shows the mole-fraction histories averaged over 100 series of runs. The solid lines with filled circles represent the results computed with the SWS, and the dash-dotted lines with open squares represent the results from conventional DSMC.}
  \label{fig:mole_frac_rate_reaction}
\end{figure*}

To confirm that the SWS preserved the chemical reaction rates of the conventional DSMC, the mole-fraction histories of the two single-cell cases used above were compared between the SWS and the conventional DSMC methods over longer times, as shown in Fig.~\ref{fig:mole_frac_rate_reaction}. The plots were averaged over 100 independent runs. 
For the dissociation case, the two mole-fraction histories agree to within approximately $1\%$ up to $1\times 10^{-6}\,\mathrm{s}$, by which point the mole fraction of the weighted species N reaches 0.36. Deviation only develops after the weighted species becomes a major component of the mixture, reaching a difference of $+11\%$ in $X_{\mathrm{N}}$ at $3\times 10^{-6}\,\mathrm{s}$, where $X_{\mathrm{N}}$ is approximately $0.5$. It should be noted that this value exceeds the mole fraction threshold of 0.1 recommended for the application of the SWS~\cite{boyd_conservative_1996}. 
The deviation scales with the mole fraction of the weighted species, which can be explained as follows. As the dissociation reaction progresses, the mole fraction of the weighted species N increases, while the major species are consumed and decrease. Due to the correction in Eq.~\eqref{eq:DSMC_collision_probability_maxfilter}, the trace species also experience collisions in proportion to their increased mole fraction, while the consumption of $\mathrm{N_2}$ reduces the collision frequency between the major species ($\mathrm{N}_2$-$\mathrm{N}_2$). Consequently, the accumulated energy loss expressed by Eq.~\eqref{eq:split_merge_E_loss} is added to the collisions of this major pair. This increases the kinetic energy of the pair to a level sufficient for a chemical reaction to be accepted, ultimately causing the reaction to proceed excessively compared to the physical reaction rate. 

For the exchange case, conversely, the mole fraction of the trace species does not increase; equilibrium is reached within $1\times 10^{-6}\,\mathrm{s}$, and the results of the two methods showed good agreement. In the reentry application discussed in Sec.~\ref{sec:case_model}, the mole fraction of each weighted species remains below $10^{-3}$, far smaller than in the verification cases above, so the reaction rates are preserved.

\clearpage

\section{Models and analysis cases of the OREX}
\label{sec:case_model}

The analysis throughout this paper deals with 11 chemical species (N$_{2}$, O$_{2}$, NO, N, O, N$^{+}$, O$^{+}$, NO$^{+}$, N$_{2}^{+}$, O$_{2}^{+}$, e$^{-}$). The reaction rate $k_\mathrm{r}$ for the TCE model, given by the modified Arrhenius equation at the molecular level, is expressed as:
\begin{equation}
\label{eq:reaction_rate}
  k_\mathrm{r} = A {T_{\mathrm{c}}}^{\eta}
  \exp\left(-\frac{\epsilon_{\mathrm{a}}}{k_\mathrm{B} T_{\mathrm{c}}}\right),
\end{equation}
where $A$ is the pre-exponential factor, $\eta$ is the temperature exponent, $T_{\mathrm{c}}$ is the controlling temperature, $\epsilon_{\mathrm{a}}$ is the activation energy, and $k_\mathrm{B}$ is the Boltzmann constant. The coefficients for each reaction are shown in Table~\ref{tab:rate_parameters_mole}. The reaction rate coefficients used are based on the adjusted Park model of Parent et al.~\cite{parent_electron_2022, morimoto_effect_2025,bose_thermal_1997, bose_thermal_1996}, which was validated against flight experiment data in a Navier--Stokes-based multi-temperature framework. In that framework, $T_{\mathrm{c}}$ is the electron temperature $T_\mathrm{e}$ for reactions involving electrons as reactants and the translational temperature $T_\mathrm{tr}$ for all other reactions.

For a collision between species A and B with masses $m_\mathrm{A}$ and $m_\mathrm{B}$, the TCE model gives the reaction probability as~\cite{bird_molecular_1994}
\begin{equation}
\label{eq:TCE_prob}
  P_{\mathrm{r}} =
  \begin{cases}
  \dfrac{\pi^{1/2} A T_\mathrm{ref}^{\eta}}{2 \sigma_{T,\mathrm{ref}} \left(k_\mathrm{B} T_\mathrm{ref}\right)^{\eta-1+\omega}}
  \dfrac{\Gamma\left(\bar{\xi}+{5/2}-\omega\right)}{\Gamma\left(\bar{\xi}+\eta+{3/2}\right)}
  \left(\dfrac{m_{\mathrm{r}}}{2 k_\mathrm{B} T_\mathrm{ref}}\right)^{1/2}
  \dfrac{(\epsilon_{\mathrm{c}} - \epsilon_{\mathrm{a}})^{\eta+\bar{\xi}+1/2}}{\epsilon_{\mathrm{c}}^{\bar{\xi}+3/2-\omega}}
  & (\epsilon_{\mathrm{c}} > \epsilon_{\mathrm{a}}), \\[6pt]
  0 & (\epsilon_{\mathrm{c}} \leq \epsilon_{\mathrm{a}}), 
  \end{cases}
\end{equation}
where $m_{\mathrm{r}} = m_\mathrm{A}m_\mathrm{B}/(m_\mathrm{A}+m_\mathrm{B})$ is the reduced mass, $T_\mathrm{ref}$ and $\sigma_{T,\mathrm{ref}}$ are the VSS reference temperature and total collision cross section, $\omega$ is the viscosity index, and $\bar{\xi}$ is the effective number of internal degrees of freedom contributing to the reaction. The total collision energy $\epsilon_{\mathrm{c}}$ is defined as the sum of the translational collision energy $\epsilon_{\mathrm{tr}} = \tfrac{1}{2} m_{\mathrm{r}} g_\mathrm{r}^2$ and the internal energy $\epsilon_{\mathrm{int}}$, i.e., $\epsilon_{\mathrm{c}} = \tfrac{1}{2} m_{\mathrm{r}} g_\mathrm{r}^2 + \epsilon_\mathrm{int}$. For reactions involving electrons, the electron mass $m_\mathrm{e}$ is much smaller than that of any heavy-particle collision partner $m_\mathrm{h}$ and the electron speed $v_\mathrm{e}$ is much greater than the heavy-particle speed $v_\mathrm{h}$, i.e., $m_\mathrm{e} \ll m_\mathrm{h}$ and $v_\mathrm{e} \gg v_\mathrm{h}$, giving $m_{\mathrm{r}} \approx m_\mathrm{e}$, $g_\mathrm{r} \approx v_{\mathrm{e}}$, and $\epsilon_{\mathrm{tr}} = \tfrac{1}{2} m_{\mathrm{r}} g_\mathrm{r}^2 \approx \tfrac{1}{2} m_{\mathrm{e}} v_\mathrm{e}^2$~\cite{li_modeling_2011}. Since $\epsilon_{\mathrm{tr}}$ is determined by the electron speed, which is characterized by the electron temperature $T_{\mathrm{e}}$ under the Maxwell--Boltzmann distribution, the TCE computation is effectively consistent with setting $T_{\mathrm{c}} = T_{\mathrm{e}}$ for electron-involved reactions in the multi-temperature framework of Parent et al.~\cite{parent_electron_2022, morimoto_effect_2025}. The same treatment was employed in previous DSMC studies of ionized hypersonic flows~\cite{ozawa_development_2008}.

The three-body recombination reactions are not considered. Some of the backward reaction rates were not given explicitly and were determined by dividing the forward reaction rates by the equilibrium constant, which was computed using the polynomial fits~\cite{park_nonequilibrium_1990, mcbride_nasa_2002, boyd_computation_1994, boyd_modeling_2007}. 
The electron impact ionization reactions ($\mathrm{X} + \mathrm{e^-} \rightarrow \mathrm{X^+} + 2\mathrm{e^-}$ for $\mathrm{X} = \mathrm{N}$, $\mathrm{O}$, $\mathrm{N_2}$, $\mathrm{O_2}$, and $\mathrm{NO}$) were taken into account; however, the reactions did not occur throughout the computations. 

In addition to the baseline chemical reaction rate set, computations were performed with the forward associative-ionization rates ($\mathrm{N}+\mathrm{O}\rightarrow\mathrm{NO^+}+\mathrm{e^-}$, $\mathrm{O}+\mathrm{O}\rightarrow\mathrm{O_2^+}+\mathrm{e^-}$, and $\mathrm{N}+\mathrm{N}\rightarrow\mathrm{N_2^+}+\mathrm{e^-}$) replaced by the recommended values of Aiken et al.~\cite{aiken_review_2025}. To quantify the rate uncertainty, the $\mathrm{N}+\mathrm{O}$ rate, the dominant electron-producing reaction at the present flight velocity~\cite{aiken_review_2025}, was additionally set to the upper and lower bounds of its recommended uncertainty interval of $4.0$ and $0.3$ times the recommended value, with the other two rates kept at their recommended values. The backward rates and all other reactions were unchanged from the baseline. The adopted coefficients are listed in Table~\ref{tab:aiken_rates}. Reactions 6a$'$, 7a$'$, and 8a$'$ replace reactions 6a, 7a, and 8a in Table~\ref{tab:rate_parameters_mole}, with the primed numbers used throughout to indicate this replacement.

For translational to rotational relaxation of diatomic molecules, we adopted the semi-empirical formulation of the Parker model~\cite{parker_rotational_1959}. For translational to vibrational relaxation, we adopted the Millikan-White model~\cite{millikan_systematics_1963} with the high temperature correction proposed by Park~\cite{park_review_1993}. 

The OREX flight conditions analyzed in this paper are those used in previous research by Moss et al.~\cite{moss_dsmc_1996} using DSMC for the evaluation of heat flux and aerodynamic characteristics. Four flow conditions shown in Table~\ref{tab:flow_conditions} were used for the analysis. In the two low-altitude cases, the ion current was saturated and within the measurable range of electron number density.

\begin{figure}[hbtp]
  \centering
  \includegraphics[width=0.65\textwidth]{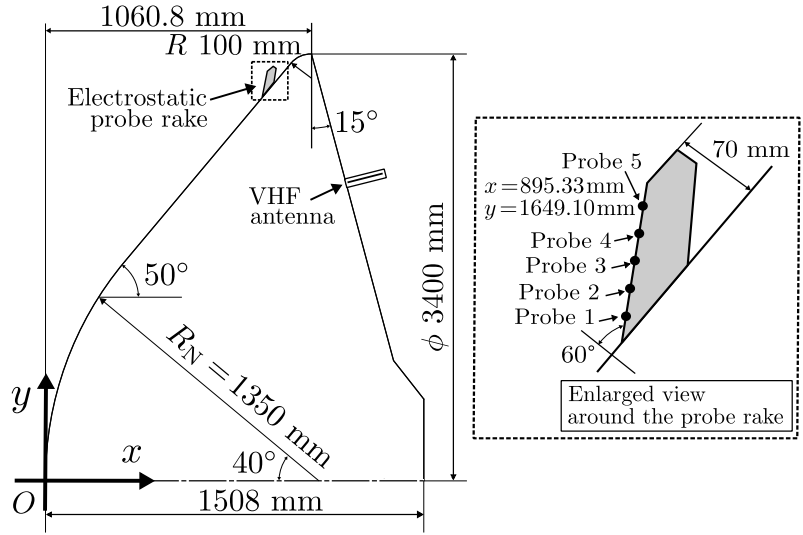}
  \caption{Schematic of OREX vehicle configuration and electrostatic probe rake geometry. 
The enlarged view shows the arrangement of Probes 1--5 on the rake.}
  \label{fig:orex_configuration}
  \vspace{-5mm}
\end{figure}

Figure~\ref{fig:orex_configuration} shows a schematic of the OREX vehicle configuration and the electrostatic probe rake mounted on the body, with Probes 1--5 arranged along the rake. In this study, a two-dimensional axisymmetric coordinate system is defined with its origin at the vehicle nose tip, where the $x$-axis points downstream along the central axis and the $y$-axis represents the radial distance from the axis of symmetry.

In the OREX simulation cases using the SWS, the detailed computational setup is summarized in Table~\ref{tab:computational_conditions}. The species weight for all neutral species is set to $w_i = 1.0$. The species weights $w_i$ for ions and electrons are chosen identically to satisfy charge conservation. Specifically, for Cases (a) and (b), we set $w_i = 1.0\times10^{-3}$ for all ion species and electrons; for Cases (c) and (d), we use $w_i = 1.0\times10^{-2}$. The global weighting factor is $F_N = 5.0\times10^{13}$ in Cases (a) and (b), and $F_N = 1.0\times10^{14}$ in Cases (c) and (d). Computations of conventional DSMC (without SWS) were performed to establish a verification baseline. Under the assumption of a small angle of attack, two-dimensional axisymmetric simulations were performed. The computational domain was set to 5 m in the radial direction from the axis, 10 m downstream from the nose in the axial direction, and 2 m upstream from the nose in the axial direction (3 m only for Case (a), 105.0~km) to capture thick shock structures. A uniform Cartesian grid with an initial cell size of $\Delta x = \Delta y = 4.00 \times 10^{-2} \,\mathrm{m}$ was used. Adaptive mesh refinement with up to five levels was applied. In all cases, the vehicle wall was treated as a fully diffuse surface with full thermal accommodation. Two wall treatments were used for the charged species. One was a non-catalytic wall and the other was a wall with surface ion recombination, where every ion that collides with the wall is neutralized and re-emitted as a neutral species, while the attached electron is removed under the ambipolar approximation. Ion recombination at the surface is considered to be a more accurate wall boundary condition for weakly ionized flows \cite{park_nonequilibrium_1990, candler_computation_1991, shoev_dsmc_2020}. The wall temperature was set to 331.8 K, 401.5 K, 485.2 K, and 586.1 K for Cases (a) through (d), respectively~\cite{moss_dsmc_1996}. The time step $\Delta t$ was set to be smaller than the mean collision time for the heavy species. Macroscopic flow field quantities were time-averaged over 100000 time steps, sampling every 10 time steps, after reaching a steady state. Sampling was started after the total number of simulated particles had reached a plateau, and this number varied by less than $0.3\%$ of its mean over the averaging duration in all cases. It was confirmed through the assessment described in Appendix~\ref{sec:appendix_convergence} that the adopted values of $F_N$, $w_i$, and the averaging duration are appropriate to obtain converged fields.

The total number of simulated particles is shown in Table~\ref{tab:computational_conditions}. The SWS cases have a slight increase relative to the conventional DSMC cases due to the production of additional minor species particles from chemical reactions.

\clearpage

\begin{table}[htbp]
\centering
\caption{Reaction rate coefficients for all reactions.}
\label{tab:rate_parameters_mole}
\makeatletter
\@ifclasswith{revtex4-2}{preprint}{%
  \renewcommand{\arraystretch}{0.65}  
}{%
  \renewcommand{\arraystretch}{1.0}   
}
\makeatother
\begin{tabular}{llccc}
\toprule
No. & Reaction & $A$ (m$^3$/molecule/s) & $\eta$ & $\epsilon_{\mathrm{a}} / k_\mathrm{B}$ (K) \\
\midrule
1M$_1$  & $\mathrm{N_2} + \mathrm{M_1} \rightarrow 2\mathrm{N} + \mathrm{M_1}$           & $4.982 \times 10^{-8}$  & $-1.6$ & $113200$ \\
1M$_2$  & $\mathrm{N_2} + \mathrm{M_2} \rightarrow 2\mathrm{N} + \mathrm{M_2}$           & $1.162 \times 10^{-8}$  & $-1.6$ & $113200$ \\
2M$_1$  & $\mathrm{O_2} + \mathrm{M_1} \rightarrow 2\mathrm{O} + \mathrm{M_1}$           & $1.661 \times 10^{-8}$  & $-1.5$ & $59500$  \\
2M$_2$  & $\mathrm{O_2} + \mathrm{M_2} \rightarrow 2\mathrm{O} + \mathrm{M_2}$           & $3.321 \times 10^{-9}$  & $-1.5$ & $59500$  \\
3M$_3$  & $\mathrm{NO} + \mathrm{M_3} \rightarrow \mathrm{N} + \mathrm{O} + \mathrm{M_3}$ & $1.827 \times 10^{-13}$ & $0$    & $75500$  \\
3M$_4$  & $\mathrm{NO} + \mathrm{M_4} \rightarrow \mathrm{N} + \mathrm{O} + \mathrm{M_4}$ & $8.303 \times 10^{-15}$ & $0$    & $75500$  \\
4a  & $\mathrm{O} + \mathrm{NO} \rightarrow \mathrm{N} + \mathrm{O_2}$ & {$5.048 \times 10^{-21}$} & {$1.084$} & {$24520$} \\
4b  & $\mathrm{N} + \mathrm{O_2} \rightarrow \mathrm{O} + \mathrm{NO}$ & {$4.139 \times 10^{-21}$} & {$1.179$} & {$4005$}  \\
5a  & $\mathrm{O} + \mathrm{N_2} \rightarrow \mathrm{N} + \mathrm{NO}$ & {$9.449 \times 10^{-18}$} & {$0.42$}  & {$42938$} \\
5b  & $\mathrm{N} + \mathrm{NO} \rightarrow \mathrm{O} + \mathrm{N_2}$ & {$4.394 \times 10^{-19}$} & {$0.513$} & {$0$}     \\
6a  & $\mathrm{N} + \mathrm{O} \rightarrow \mathrm{NO^+} + \mathrm{e^-}$          & $8.801 \times 10^{-18}$ & $0$    & $32000$  \\
6b  & $\mathrm{NO^+} + \mathrm{e^-} \rightarrow \mathrm{N} + \mathrm{O}$          & $9.748 \times 10^{-13}$ & $-0.300$ & $100$ \\
7a  & $\mathrm{O} + \mathrm{O} \rightarrow \mathrm{O_2^+} + \mathrm{e^-}$         & $1.827 \times 10^{-17}$ & $0$    & $81200$  \\
7b  & $\mathrm{O_2^+} + \mathrm{e^-} \rightarrow \mathrm{O} + \mathrm{O}$         & $2.524 \times 10^{-12}$ & $-0.341$ & $881$ \\
8a  & $\mathrm{N} + \mathrm{N} \rightarrow \mathrm{N_2^+} + \mathrm{e^-}$         & $3.321 \times 10^{-17}$ & $0$    & $67700$  \\
8b  & $\mathrm{N_2^+} + \mathrm{e^-} \rightarrow \mathrm{N} + \mathrm{N}$         & $7.722 \times 10^{-13}$ & $-0.249$ & $7$   \\
9a  & $\mathrm{NO^+} + \mathrm{O} \rightarrow \mathrm{N^+} + \mathrm{O_2}$        & $1.661 \times 10^{-18}$ & $0.5$  & $77200$  \\
9b  & $\mathrm{N^+} + \mathrm{O_2} \rightarrow \mathrm{NO^+} + \mathrm{O}$        & $3.040 \times 10^{-18}$ & $0.29$ & $0$      \\
10a & $\mathrm{N^+} + \mathrm{N_2} \rightarrow \mathrm{N_2^+} + \mathrm{N}$       & $1.661 \times 10^{-18}$ & $0.5$  & {$12200$}  \\
10b & $\mathrm{N} + \mathrm{N_2^+} \rightarrow \mathrm{N_2} + \mathrm{N^+}$       & $2.343 \times 10^{-14}$ & $-0.61$ & $0$     \\
11a & $\mathrm{O_2^+} + \mathrm{N} \rightarrow \mathrm{N^+} + \mathrm{O_2}$       & $1.445 \times 10^{-16}$ & $0.14$ & $28600$  \\
11b & $\mathrm{N^+} + \mathrm{O_2} \rightarrow \mathrm{O_2^+} + \mathrm{N}$       & $2.189 \times 10^{-21}$ & $1.227$ & $0$     \\
12a & $\mathrm{O^+} + \mathrm{NO} \rightarrow \mathrm{N^+} + \mathrm{O_2}$        & $2.325 \times 10^{-25}$ & $1.9$  & $26600$  \\
12b & $\mathrm{N^+} + \mathrm{O_2} \rightarrow \mathrm{O^+} + \mathrm{NO}$        & $2.443 \times 10^{-26}$ & $2.102$ & $0$     \\
13a & $\mathrm{O_2^+} + \mathrm{N_2} \rightarrow \mathrm{N_2^+} + \mathrm{O_2}$   & $1.644 \times 10^{-17}$ & $0$    & $40700$  \\
13b & $\mathrm{N_2^+} + \mathrm{O_2} \rightarrow \mathrm{O_2^+} + \mathrm{N_2}$   & $4.589 \times 10^{-18}$ & $-0.037$ & $0$   \\
14a & $\mathrm{O_2^+} + \mathrm{O} \rightarrow \mathrm{O^+} + \mathrm{O_2}$       & $6.642 \times 10^{-18}$ & $-0.09$ & {$18000$} \\
14b & $\mathrm{O^+} + \mathrm{O_2} \rightarrow \mathrm{O_2^+} + \mathrm{O}$       & $4.993 \times 10^{-18}$ & $-0.004$ & $0$   \\
15a & $\mathrm{NO^+} + \mathrm{N} \rightarrow \mathrm{O^+} + \mathrm{N_2}$        & $5.646 \times 10^{-17}$ & $-1.08$ & $12800$ \\
15b & $\mathrm{O^+} + \mathrm{N_2} \rightarrow \mathrm{NO^+} + \mathrm{N}$        & $3.970 \times 10^{-18}$ & $-0.71$ & $0$     \\
16a & $\mathrm{NO^+} + \mathrm{O_2} \rightarrow \mathrm{O_2^+} + \mathrm{NO}$     & $3.985 \times 10^{-17}$ & $0.41$ & {$32600$}  \\
16b & $\mathrm{O_2^+} + \mathrm{NO} \rightarrow \mathrm{NO^+} + \mathrm{O_2}$     & $6.195 \times 10^{-16}$ & $-0.05$ & $0$     \\
17a & $\mathrm{NO^+} + \mathrm{O} \rightarrow \mathrm{O_2^+} + \mathrm{N}$        & $1.195 \times 10^{-17}$ & $0.29$ & $48600$  \\
17b & $\mathrm{O_2^+} + \mathrm{N} \rightarrow \mathrm{NO^+} + \mathrm{O}$        & $8.918 \times 10^{-13}$ & $-0.969$ & $0$   \\
18a & $\mathrm{O^+} + \mathrm{N_2} \rightarrow \mathrm{N_2^+} + \mathrm{O}$       & $1.511 \times 10^{-18}$ & $0.36$ & {$22800$}  \\
18b & $\mathrm{N_2^+} + \mathrm{O} \rightarrow \mathrm{O^+} + \mathrm{N_2}$       & $1.978 \times 10^{-18}$ & $0.109$ & $0$     \\
19a & $\mathrm{NO^+} + \mathrm{N} \rightarrow \mathrm{N_2^+} + \mathrm{O}$        & $1.196 \times 10^{-16}$ & $0$    & $35500$  \\
19b & $\mathrm{N_2^+} + \mathrm{O} \rightarrow \mathrm{NO^+} + \mathrm{N}$        & $1.744 \times 10^{-18}$ & $0.302$ & $0$     \\
\bottomrule
\multicolumn{5}{l}{M$_1$ = N, O; \quad M$_2$ = N$_2$, O$_2$, NO; \quad M$_3$ = N, O, NO; \quad M$_4$ = N$_2$, O$_2$.} \\
\end{tabular}
\end{table}
\clearpage

\begin{table}[htbp]
\centering
\caption{Forward associative-ionization rate coefficients recommended by Aiken et al.~\cite{aiken_review_2025}. The upper and lower bounds of the $\mathrm{N} + \mathrm{O}$ rate correspond to $A = 7.240 \times 10^{-18}$ and $A = 5.430 \times 10^{-19}$.}
\label{tab:aiken_rates}
\makeatletter
\@ifclasswith{revtex4-2}{preprint}{%
  \renewcommand{\arraystretch}{0.65}  
}{%
  \renewcommand{\arraystretch}{1.0}   
}
\makeatother
\begin{tabular}{llccc}
\toprule
No. & Reaction & $A$ (m$^3$/molecule/s) & $\eta$ & $\epsilon_{\mathrm{a}} / k_\mathrm{B}$ (K) \\
\midrule
6a$'$ & $\mathrm{N} + \mathrm{O} \rightarrow \mathrm{NO^+} + \mathrm{e^-}$  & $1.810 \times 10^{-18}$ & $0.15$ & $34586$ \\
7a$'$ & $\mathrm{O} + \mathrm{O} \rightarrow \mathrm{O_2^+} + \mathrm{e^-}$ & $3.022 \times 10^{-20}$ & $0.68$ & $80640$ \\
8a$'$ & $\mathrm{N} + \mathrm{N} \rightarrow \mathrm{N_2^+} + \mathrm{e^-}$ & $5.795 \times 10^{-19}$ & $0.51$ & $71211$ \\
\bottomrule
\end{tabular}
\end{table}

\begin{table*}[htbp]
  \centering
  \caption{Freestream conditions for the OREX flight simulation}
  \label{tab:flow_conditions}
  \begin{tabular}{ccccccc}
  \hline
  Case
  & Altitude & Velocity & Temperature & Number Density & Knudsen Number & Mach Number\\
  & (km) & (m/s) & (K) & ($\mathrm{m}^{-3}$) & & \\
  \hline
    (a) & $105.0$ & $7451$ & $211.0$ & $5.05 \times 10^{18}$ & $7.2 \times 10^{-2}$ & $25.1$ \\
    (b) & $101.1$ & $7455$ & $196.9$ & $1.03 \times 10^{19}$ & $3.5 \times 10^{-2}$ & $26.5$ \\
    (c) & $96.8 $ & $7456$ & $190.3$ & $1.98 \times 10^{19}$ & $1.8 \times 10^{-2}$ & $26.7$ \\
    (d) & $92.8 $ & $7454$ & $188.3$ & $4.08 \times 10^{19}$ & $8.6 \times 10^{-3}$ & $27.0$ \\
  \hline
  \end{tabular}
  \vspace{-5mm}
\end{table*}

\begin{table*}[!htbp]
  \centering
  \caption{Computational setup for the OREX flight simulation cases.}
  \label{tab:computational_conditions}
  \begin{tabular}{cccccccc}
  \hline
  & & & & & \multicolumn{2}{c}{Total number of particles} & \\
  \cline{6-7}
  Case & Altitude & $\Delta t$ & $F_N$ & $w_i$ for ions and electrons & Conventional & SWS & \\
  & (km) & (s) & &  &(without SWS) & & \\
  \hline
    (a) & $105.0$ & $1.0 \times 10^{-6}$ & $5.0 \times 10^{13}$  & $1.0 \times 10^{-3}$ & $4.042 \times 10^{7}$ & $4.051 \times 10^{7}$ &  \\
    (b) & $101.1$ & $1.0 \times 10^{-7}$ & $5.0 \times 10^{13}$  & $1.0 \times 10^{-3}$ & $7.523 \times 10^{7}$ & $7.728 \times 10^{7}$ &  \\
    (c) & $96.8$  & $1.0 \times 10^{-7}$ & $1.0 \times 10^{14}$  & $1.0 \times 10^{-2}$ & $7.367 \times 10^{7}$ & $7.433 \times 10^{7}$ &  \\
    (d) & $92.8$  & $1.0 \times 10^{-7}$ & $1.0 \times 10^{14}$  & $1.0 \times 10^{-2}$ & $1.563 \times 10^{8}$ & $1.587 \times 10^{8}$ &  \\
  \hline
  \end{tabular}
\end{table*}

\clearpage

\section{RESULTS AND DISCUSSION}
\label{sec:results_discussion}

\subsection{Temperature comparison and shock wave analysis}
\label{sec:temperature_shock_analysis}

\begin{figure*}[htbp]
 \centering 
\begin{subfigure}[t]{0.47\textwidth}
  \centering
  \includegraphics[width=1.08\textwidth]{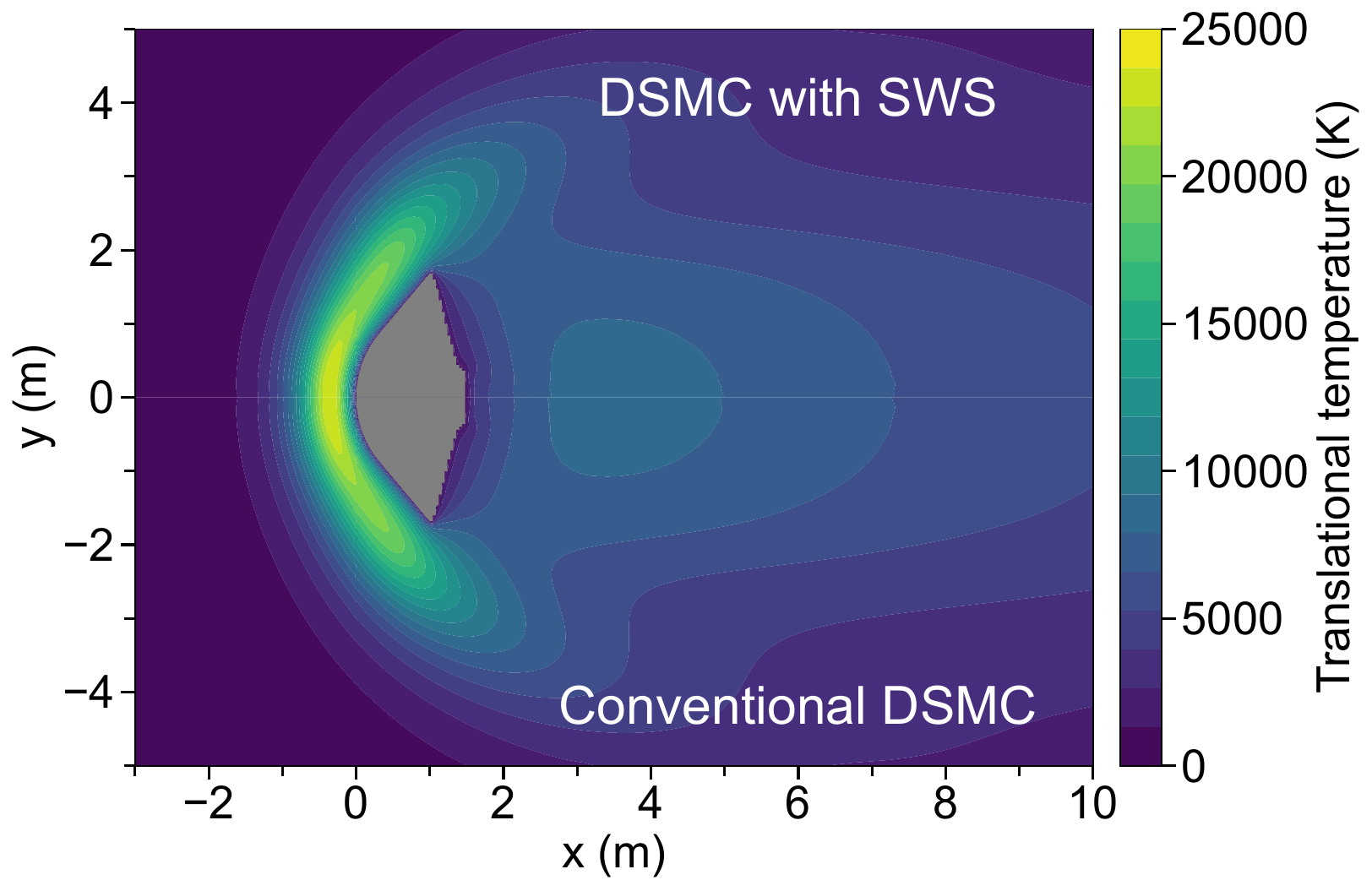}
  \caption{Altitude 105.0~km} 
  \label{temp_field_SWS_105} 
\end{subfigure}
 \hfill
 \begin{subfigure}[t]{0.47\textwidth}
   \centering
   \includegraphics[width=\textwidth]{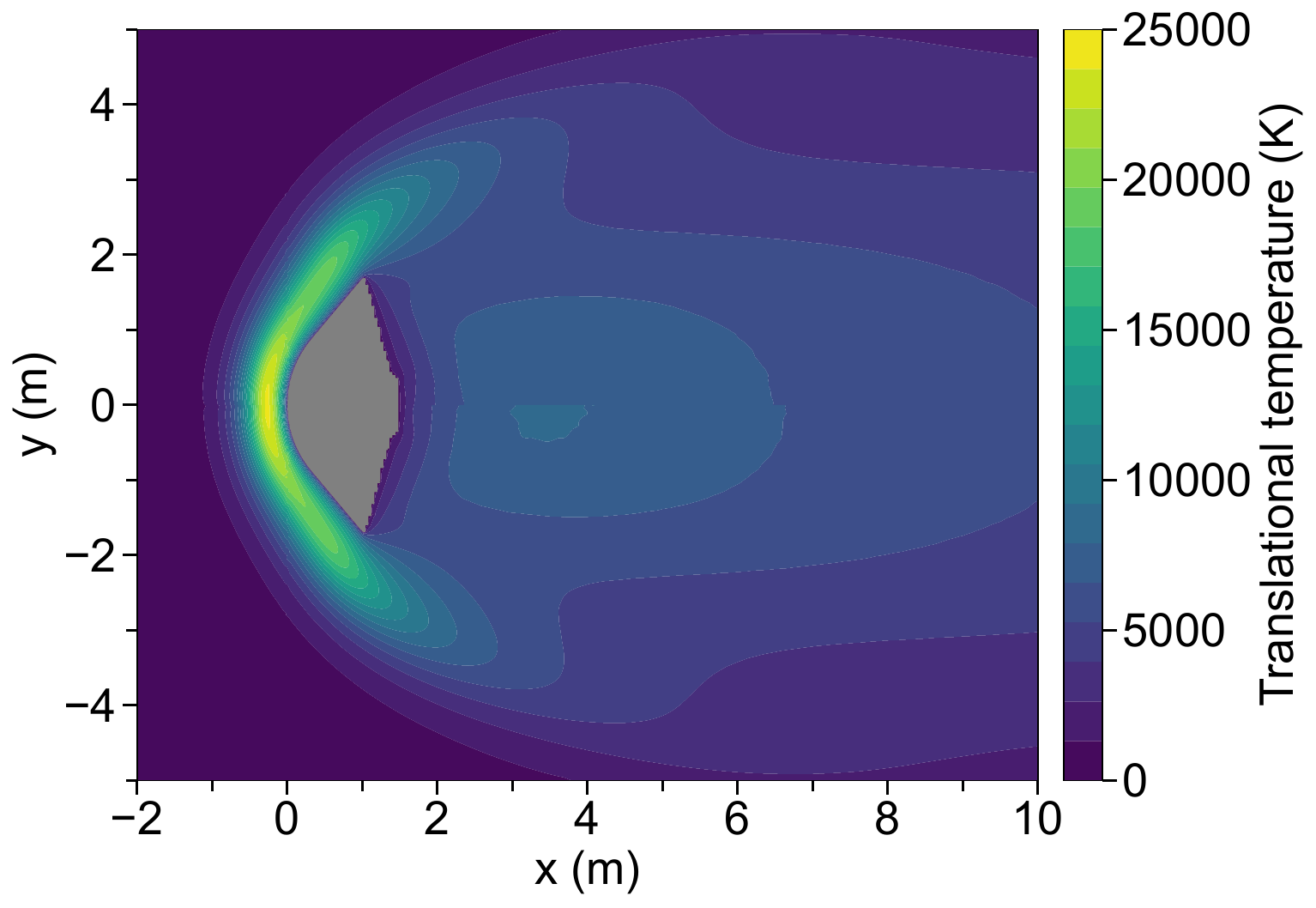}
   \caption{Altitude 101.1~km}
   \label{temp_field_SWS_101}
 \end{subfigure}
 \hfill
 \begin{subfigure}[t]{0.47\textwidth}
   \centering
   \includegraphics[width=\textwidth]{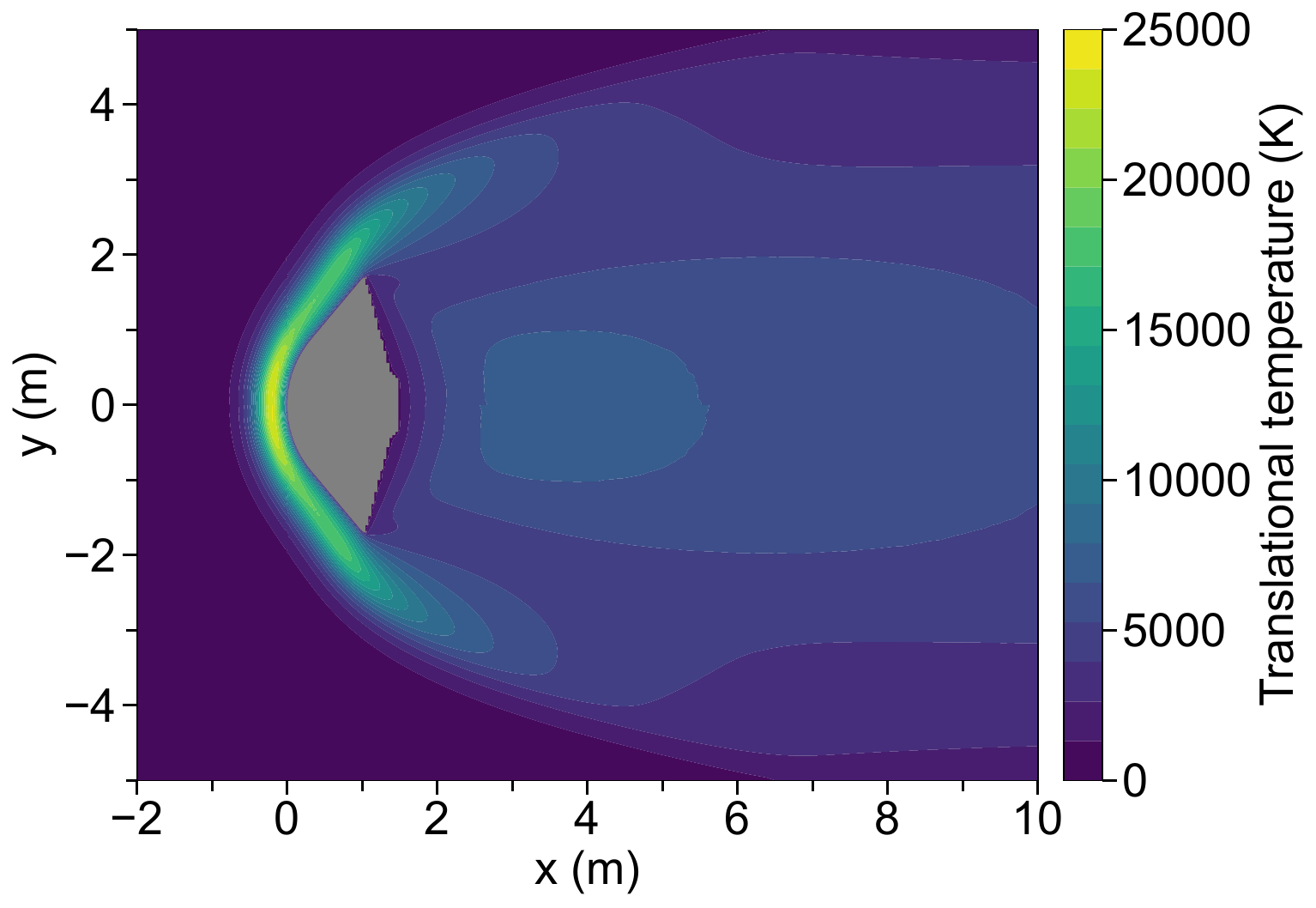}
   \caption{Altitude 96.8~km}
   \label{temp_field_SWS_96}
 \end{subfigure}
 \hfill
 \begin{subfigure}[t]{0.47\textwidth}
   \centering
   \includegraphics[width=\textwidth]{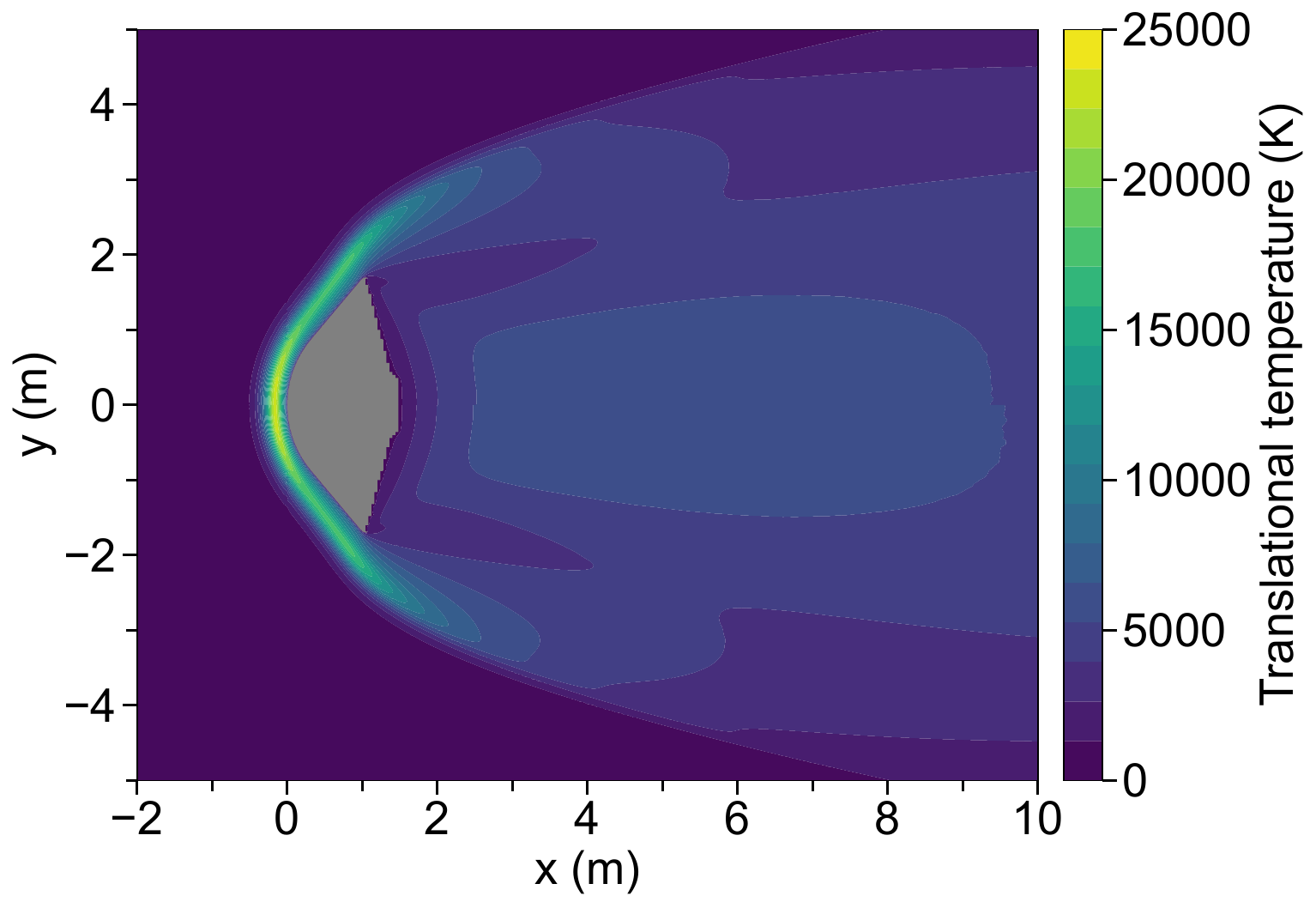}
   \caption{Altitude 92.8~km}
   \label{temp_field_SWS_92}
 \end{subfigure}
 \caption{Visualized translational temperature (K) distribution for each case: (a) Altitude 105.0~km, (b) Altitude 101.1~km, (c) Altitude 96.8~km, and (d) Altitude 92.8~km. The upper half of each figure shows the SWS results, while the lower half shows the conventional DSMC results.}
 \label{fig:temp_field_comparison_SWS}
\end{figure*}

This section evaluates the results of OREX flight simulations obtained with the present SWS and discusses the electron number density results. The influence of the physical modeling on the predicted plasma quantities is also discussed.
First, to confirm that the bulk flow was solved correctly when using the SWS, we analyze the temperature as a macroscopic variable. The verification in this section is presented for the simulations with surface ion recombination. The choice of the surface treatment had a negligible effect on the translational, rotational, and vibrational temperatures, since the ion wall boundary condition acts only on the trace charged species. Figure~\ref{fig:temp_field_comparison_SWS} shows the translational temperature distribution, comparing the SWS results and conventional DSMC results. The translational temperature of the gas mixture obtained with SWS was in agreement with the conventional DSMC results, confirming that the bulk mixture behavior was accurately reproduced. In terms of flow field characteristics, the shock wave thickness increases as the mean free path increases with increasing altitude. In other words, the more rarefied the flow is, the thicker the shock wave becomes. This observation is a known feature of shock waves in rarefied flows. The overall flow field configuration was also in good agreement between the two methods. In the wake, a region of slightly elevated temperature is observed at the rear stagnation area along the symmetry axis, as expected for hypersonic flow around a blunt body. 

\begin{figure*}[htbp]
 \centering 
 \begin{subfigure}[t]{0.49\textwidth}
    \centering
    \includegraphics[width=\textwidth]{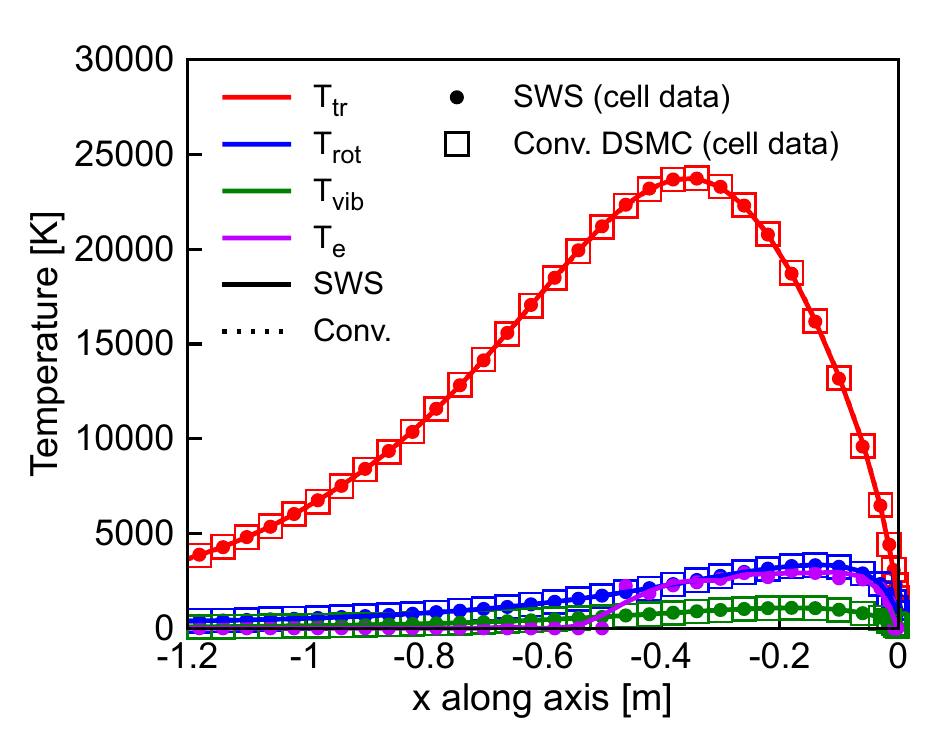}
    \caption{Altitude 105.0~km}
    \label{temp_shock_SWS_105}
 \end{subfigure} 
 \hfill
 \begin{subfigure}[t]{0.49\textwidth}
   \centering
   \includegraphics[width=\textwidth]{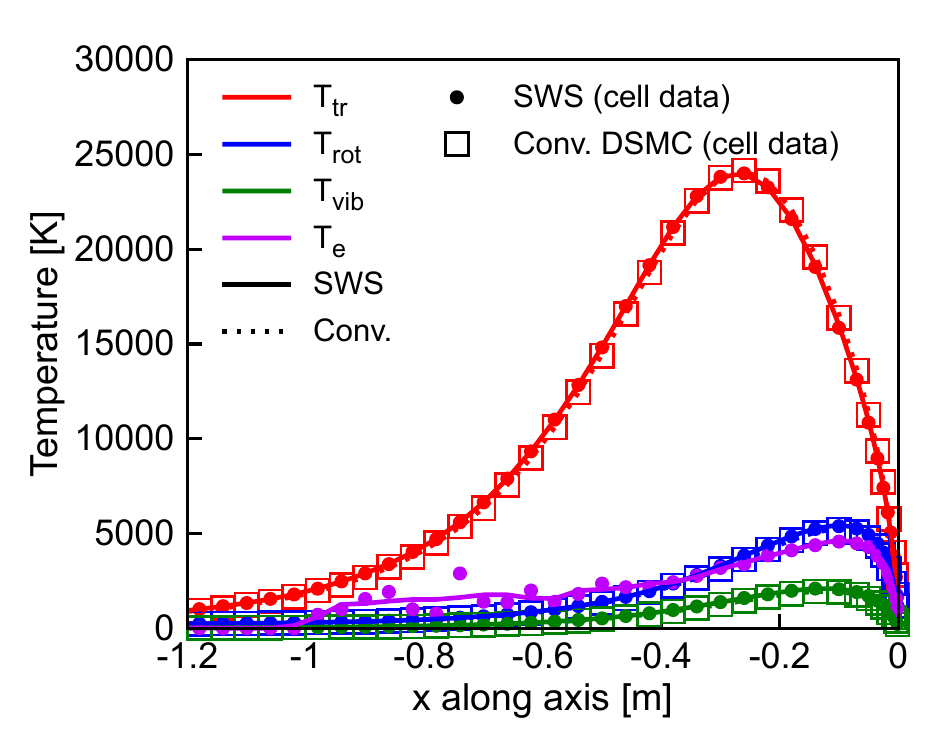}
   \caption{Altitude 101.1~km}
   \label{temp_shock_SWS_101}
 \end{subfigure}
 \hfill
 \begin{subfigure}[t]{0.49\textwidth}
   \centering
   \includegraphics[width=\textwidth]{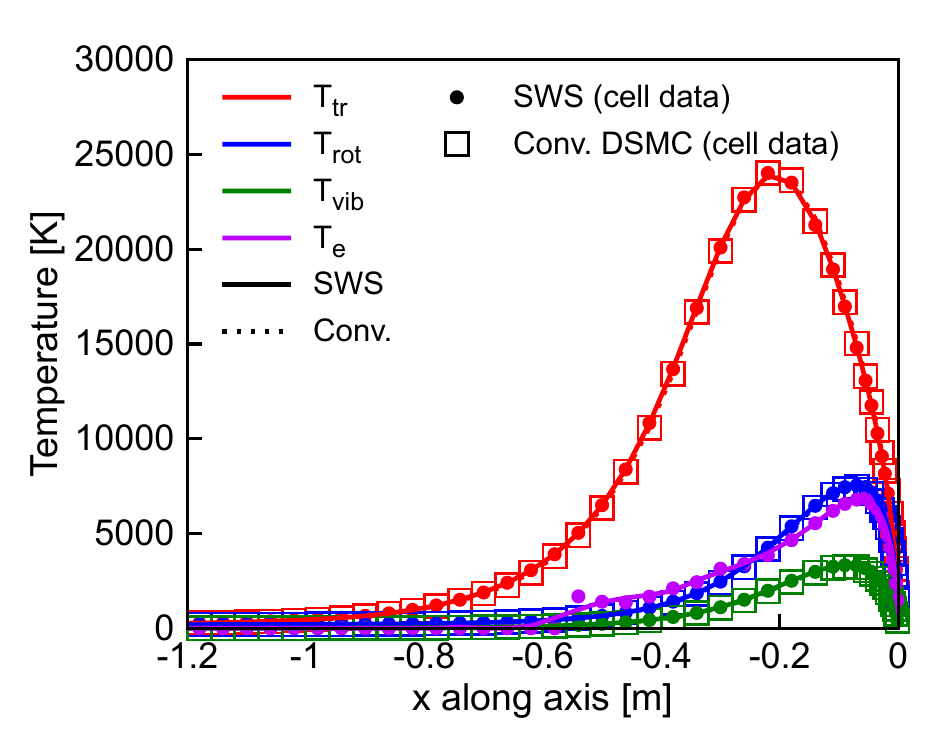}
   \caption{Altitude 96.8~km}
   \label{temp_shock_SWS_96}
 \end{subfigure}
 \hfill
 \begin{subfigure}[t]{0.49\textwidth}
   \centering
   \includegraphics[width=\textwidth]{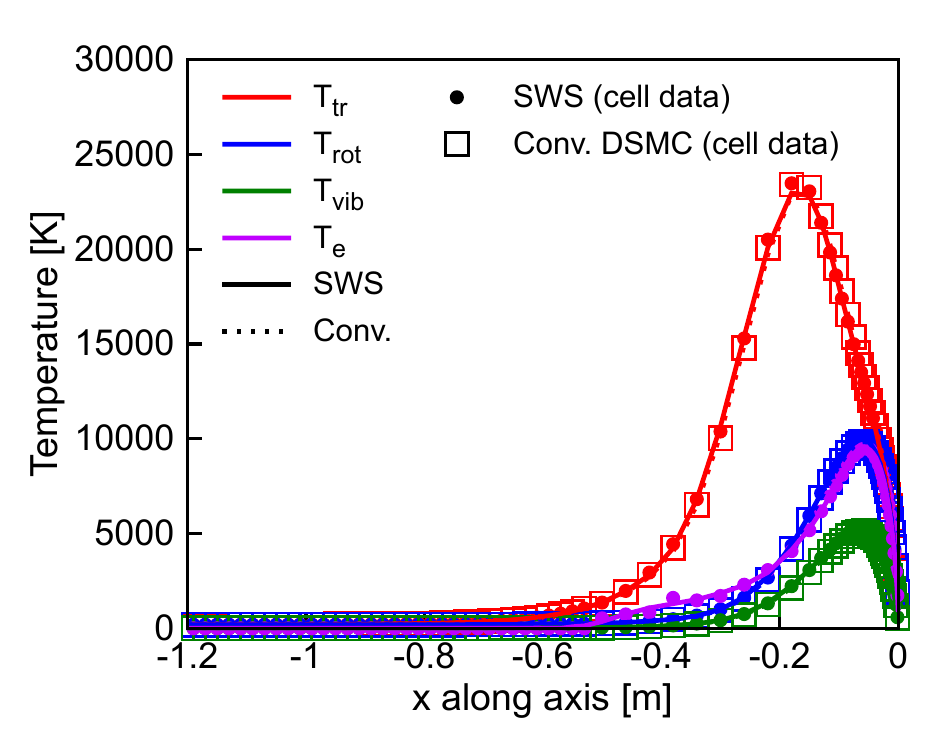}
   \caption{Altitude 92.8~km}
   \label{temp_shock_SWS_92}
 \end{subfigure}
 \caption{Translational temperature ($T_\mathrm{tr}$), vibrational temperature ($T_\mathrm{vib}$), rotational temperature ($T_\mathrm{rot}$), and electron temperature ($T_\mathrm{e}$) of the shock wave along the axis to the stagnation point. The filled circles represent SWS cell data. The open squares represent conventional DSMC cell data. The solid and dotted lines indicate the smoothed SWS and conventional DSMC results, respectively: (a) Altitude 105.0~km, (b) Altitude 101.1~km, (c) Altitude 96.8~km, and (d) Altitude 92.8~km.}
 \label{fig:temp_shock_comparison_SWS}
\end{figure*}

Figure~\ref{fig:temp_shock_comparison_SWS} shows the translational temperature $T_\mathrm{tr}$, vibrational temperature $T_\mathrm{vib}$, rotational temperature $T_\mathrm{rot}$, and electron temperature $T_\mathrm{e}$ along the stagnation streamline for each altitude case, comparing the SWS results with the conventional DSMC results as a baseline. The smoothed profiles were obtained by applying an 11-point fourth-order nonuniform Savitzky--Golay filter~\cite{gallis_accelerating_2006, gorry_general_1990} to the SWS and conventional DSMC results, where nonuniform point spacing arises from adaptive mesh refinement. The peak value of $T_\mathrm{tr}$ in the smoothed profiles differed from that in the cell data by at most $2.2\,\%$, and its location was unchanged. Note that the electron temperature $T_\mathrm{e}$ is only available from the SWS results, as the conventional DSMC has difficulty resolving the electron behavior, since electrons are trace species in the present reentry scenario at high altitude.

The cell data and the smoothed profiles of the two methods showed good overall agreement across all altitude cases, confirming that the shock structure along the stagnation streamline is correctly solved by the present SWS. 
As altitude decreases, the shock wave becomes thinner and the peak temperature location shifts closer to the vehicle nose. The peak of $T_\mathrm{tr}$ was located approximately $0.34$, $0.26$, $0.22$, and $0.18$~m ahead of the nose for Cases (a) through (d), respectively, as seen in the temperature distributions of Fig.~\ref{fig:temp_field_comparison_SWS}. 
Relative to the conventional DSMC values, the peak values of $T_\mathrm{tr}$, $T_\mathrm{vib}$, and $T_\mathrm{rot}$ obtained with the SWS differed by $0.6$, $7.9$, and $3.4\,\%$, respectively, corresponding to at most $180$~K, and the peak locations of the two methods agreed at all four altitudes. 
Strong thermal nonequilibrium was observed behind the shock wave, with internal energy mode temperatures $T_\mathrm{vib}$ and $T_\mathrm{rot}$ remaining below $T_\mathrm{tr}$ over most of the shock layer. In particular, in Cases (c) and (d), the successive rise of $T_\mathrm{tr}$, $T_\mathrm{rot}$, and finally $T_\mathrm{vib}$ behind the shock wave was observed, as a result of the characteristic relaxation of internal energy modes.

Focusing on electron temperature $T_\mathrm{e}$, the rise behind the shock wave occurs in two stages in Cases (c) and (d), a gradual increase followed by a steep rise to a peak. Similar behavior has been reported in Navier--Stokes-based studies~\cite{farbar_numerical_2013, takahashi_prediction_2014}. The electron temperature $T_\mathrm{e}$ is sampled from the electron simulated particles alone, whose number per cell remains small even with the species weighting applied. Statistical fluctuations in the $T_\mathrm{e}$ cell data were therefore still observed in Cases (a) and (b).

From the above, it was confirmed that the flow and the macroscopic values were correctly captured when using the SWS. 

\clearpage

\clearpage

\subsection{Electron number density evaluation}
\label{sec:electron_number_density}

\begin{figure*}[htbp]
\centering 
\begin{subfigure}[t]{0.48\textwidth}
  \centering
  \includegraphics[width=1.08\textwidth]{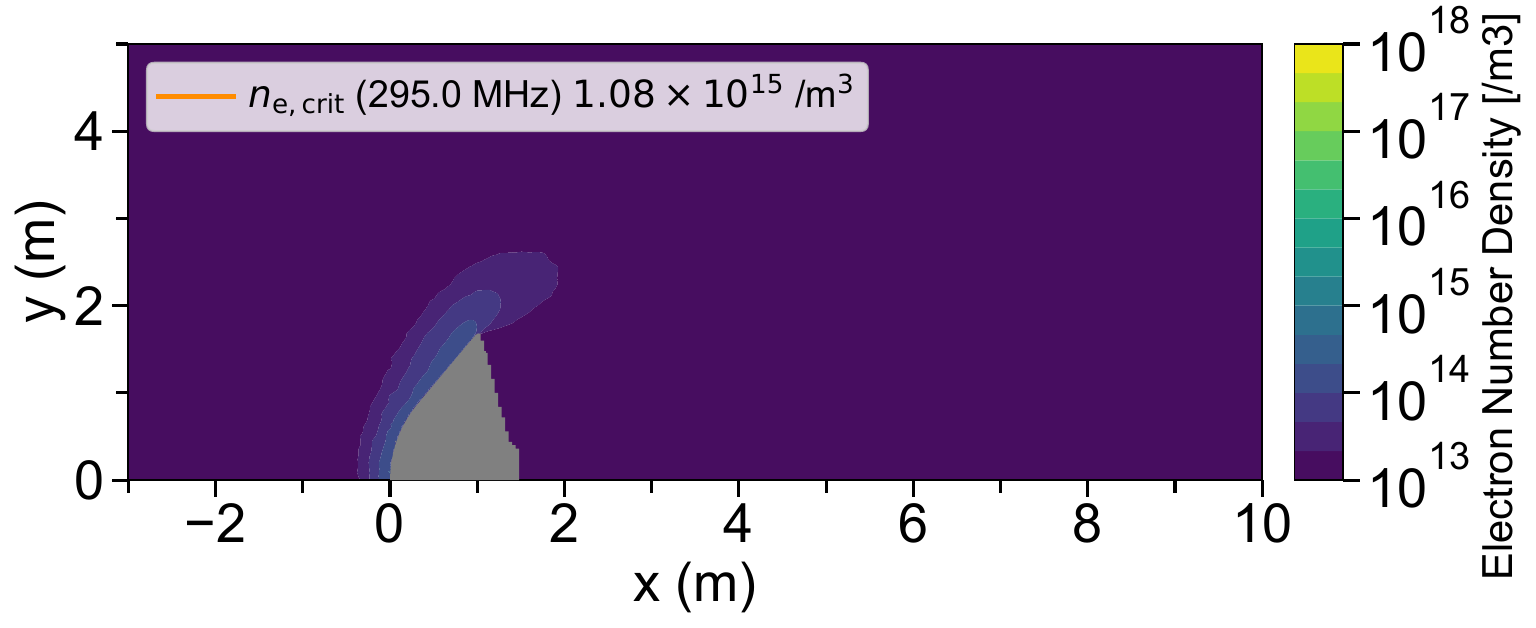}
  \caption{Altitude 105.0~km}
  \label{electron_field_SWS_105}
\end{subfigure}
\hfill
\begin{subfigure}[t]{0.48\textwidth}
   \centering
   \includegraphics[width=\textwidth]{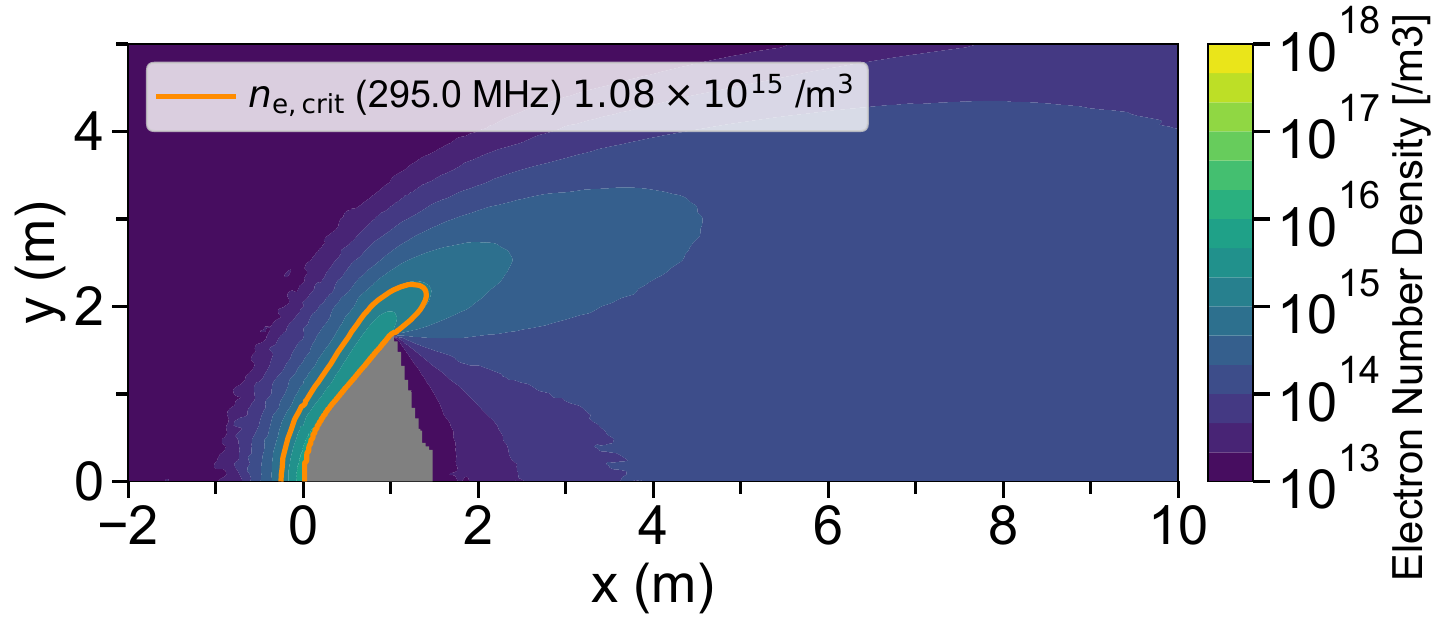}
   \caption{Altitude 101.1~km}
   \label{electron_field_SWS_101}
\end{subfigure}
\hfill
\begin{subfigure}[t]{0.48\textwidth}
   \centering
   \includegraphics[width=\textwidth]{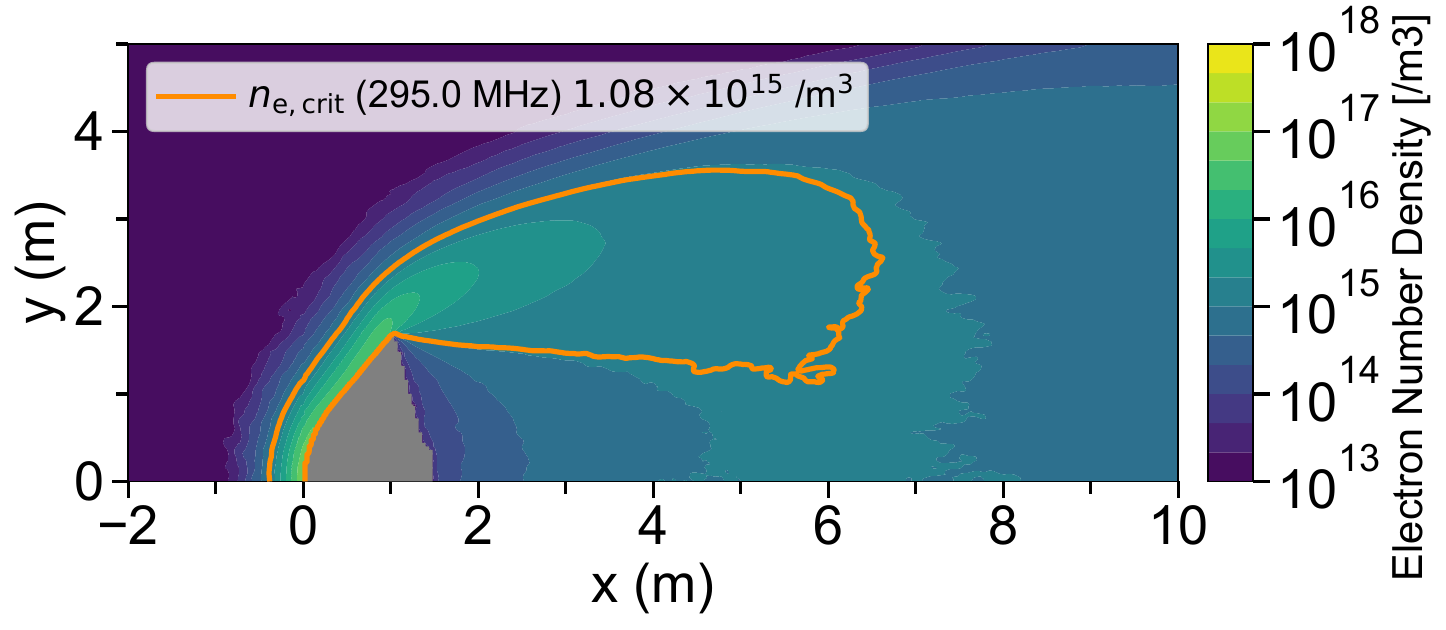}
   \caption{Altitude 96.8~km}
   \label{electron_field_SWS_96}
\end{subfigure}
\hfill
\begin{subfigure}[t]{0.48\textwidth}
   \centering
   \includegraphics[width=\textwidth]{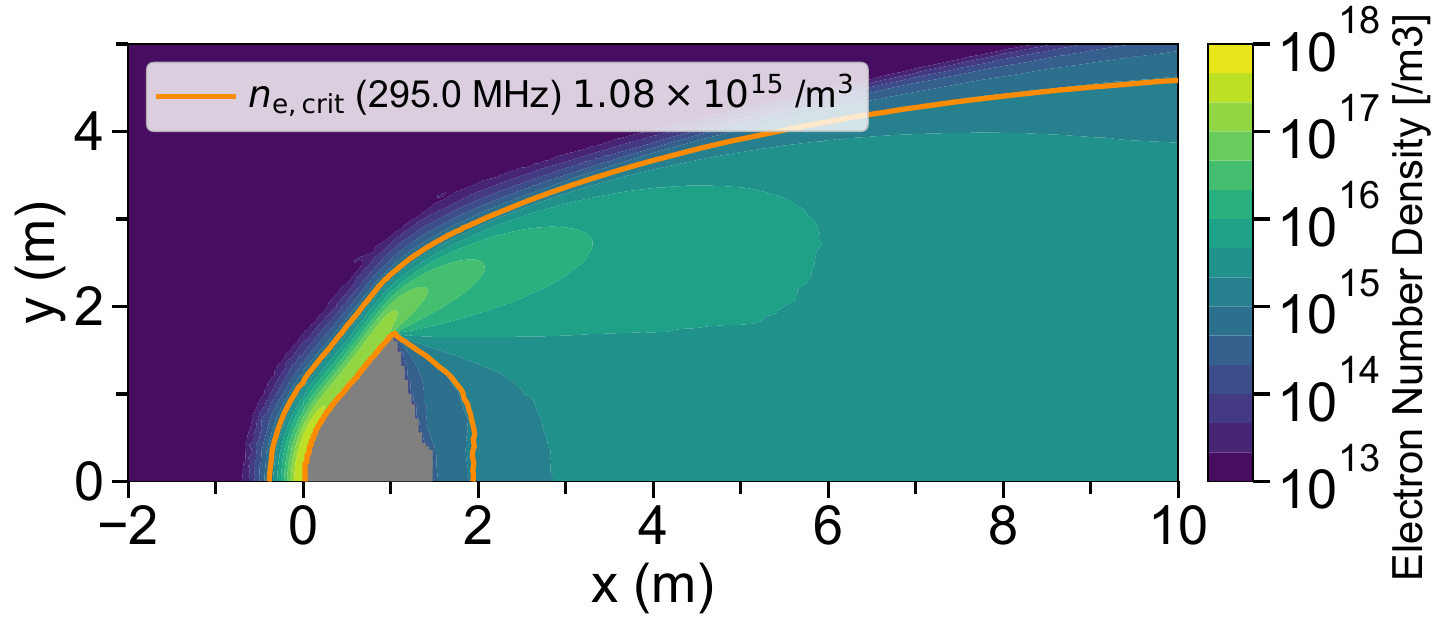}
   \caption{Altitude 92.8~km}
   \label{electron_field_SWS_92}
\end{subfigure}
\caption{Visualized electron number density ($\mathrm{m^{-3}}$) distribution obtained with the present SWS for the wall with surface ion recombination: (a) Altitude 105.0~km, (b) Altitude 101.1~km, (c) Altitude 96.8~km, and (d) Altitude 92.8~km. The orange lines indicate the isoline of the critical electron number density as defined by Eq.~\eqref{eq:ne_crit}: $n_{\mathrm{e},\mathrm{crit}} = 1.08 \times 10^{15} \, \mathrm{m^{-3}}$ for the VHF frequency of 295.0~MHz.}
\label{fig:electron_field_SWS}
\end{figure*}

Second, we present the electron number density distribution, which is the primary focus of this study and cannot be obtained with conventional DSMC or SWS without a specific reactive scheme. 
The results in this section are those obtained with the surface ion recombination described in Sec.~\ref{sec:case_model}.
Figure~\ref{fig:electron_field_SWS} shows the electron number density fields obtained with the present SWS at the four altitudes, demonstrating spatially continuous distributions within the shock layer at each altitude.

Electrons produced in the stagnation region are advected downstream into the wake region, where they undergo expansion and dissociative recombination, leading to a decrease in the electron number density on the wake side of the vehicle.
The peak electron number density on the stagnation streamline increased from approximately $3.2 \times 10^{14}\,\mathrm{m^{-3}}$ at 105.0~km to $4.4 \times 10^{17}\,\mathrm{m^{-3}}$ at 92.8~km, consistent with the increasing atmospheric number density and the consequent increase in ionization reactions at lower altitudes. The electron number density reached its maximum within the shock layer and decreased toward the vehicle wall, whereas with the non-catalytic wall it increased up to the surface.

The dense plasma region develops progressively toward the wake side with decreasing altitude. At higher altitudes, it is present only near the vehicle nose, whereas at lower altitudes it extends into the wake region downstream, indicating the overall increase in ionization in the shock layer.

Next, the relationship between the communication blackout observed during the OREX flight and the plasma distribution obtained in the present study is examined. During the OREX flight, very high frequency (VHF) signals at 295.0~MHz and 296.2~MHz and an ultra high frequency (UHF) signal at 1575.42~MHz were used~\cite{orex_1994}. The isoline of the critical electron number density ($n_{\mathrm{e},\mathrm{crit}} = 1.08 \times 10^{15}\,\mathrm{m^{-3}}$) for the VHF frequency of 295.0~MHz, as defined in Eq.~\eqref{eq:ne_crit}, is shown in Fig.~\ref{fig:electron_field_SWS}. Regions where the electron number density exceeds this threshold lead to blackout. At 101.1~km, the plasma region above the threshold $n_{\mathrm{e},\mathrm{crit}}$ reaches the vehicle shoulder from the vehicle nose. At 96.8~km, it extends further into the wake region of the vehicle, indicating that the plasma layer envelops the vehicle with decreasing altitude. This suggests that VHF blackout onset occurs at approximately 101.1~km or thereafter during the descent, which is consistent with the VHF blackout observed during the OREX flight at approximately 100~km altitude~\cite{orex_1994}. Furthermore, capturing this plasma development process allows us to analyze attenuation and diffraction using electromagnetic field analysis, enabling us to understand the process by which blackouts occur at each base station. 

To validate the quantitative accuracy of the present method, the simulated ion current $I_{\mathrm{i,s}}$ is compared with that directly measured by the electrostatic probes during the flight, because deriving the electron number density from the experimental probe data relies on temperatures obtained from the simulation, such as the electron temperature. This current is evaluated at the location of Probe 5 mounted on the rake shown in Fig.~\ref{fig:orex_configuration}, situated at an axial distance of $x = 895.33\,\mathrm{mm}$ and a radial distance of $y = 1649.10\,\mathrm{mm}$ from the vehicle nose tip~\cite{jung_numerical_2016}. The probe has a semi-cylindrical geometry with a radius of $R_\mathrm{p} = 0.1\,\mathrm{mm}$ and a length of $L_\mathrm{p} = 5\,\mathrm{mm}$, and is negatively biased at $-5\,\mathrm{V}$ relative to the electron collector.
In the present configuration, the flow velocity relative to the probe $U$ is not negligible compared to the ion thermal velocity, so the directed flow contribution should be accounted for~\cite{jones_electrostatic-probe_1972}. The ion saturation current is evaluated using the cylindrical probe model for a flowing plasma as described by Hoegy and Wharton~\cite{hoegy_current_1973} under the assumption of a collisionless sheath and orbit-limited ion collection. The current expression that incorporates both the probe bias potential and the directed flow is given by:
\begin{equation}
    I_{\mathrm{i,s}} = i_{\mathrm{r}} \frac{2}{\sqrt{\pi}} 
    \sqrt{V + s^2 + \frac{V + s^2/2}{V + s^2}} ,
    \label{eq:ion_current_HW}
\end{equation}
where $i_{\mathrm{r}} = A_\mathrm{eff} n_\mathrm{i} e (k_\mathrm{B} T_\mathrm{i} / 2 \pi m_\mathrm{i})^{1/2}$ is the random probe current, $T_\mathrm{i}$ is the ion temperature, $m_\mathrm{i}$ is the ion mass, $A_\mathrm{eff} = 2 \pi R_\mathrm{p} L_\mathrm{p}$ is the effective probe area, $V$ is the normalized potential across the sheath calculated based on Ref.~\cite{jones_electrostatic-probe_1972}, $v_\mathrm{mp} = (2 k_\mathrm{B} T_\mathrm{i} / m_\mathrm{i})^{1/2}$ is the most probable ion speed, $U$ is the flow velocity relative to the probe, and $s = U / v_\mathrm{mp}$ is the ion speed ratio.

The electron number densities obtained by the SWS at the Probe 5 position were $1.5\times10^{14}\,\mathrm{m^{-3}}$ at 105.0~km, $4.0\times10^{15}\,\mathrm{m^{-3}}$ at 101.1~km, $3.3\times10^{16}\,\mathrm{m^{-3}}$ at 96.8~km, and $1.7\times10^{17}\,\mathrm{m^{-3}}$ at 92.8~km.

The present results in the three lower-altitude cases showed agreement with the electron number densities estimated from the probe ion saturation current in Refs.~\cite{orex_1994, jung_numerical_2016}, where the electron temperatures used in those estimates were obtained from two-temperature Navier--Stokes based simulations.

\begin{figure*}[hbtp]
  \centering
  \includegraphics[width=0.55\textwidth]{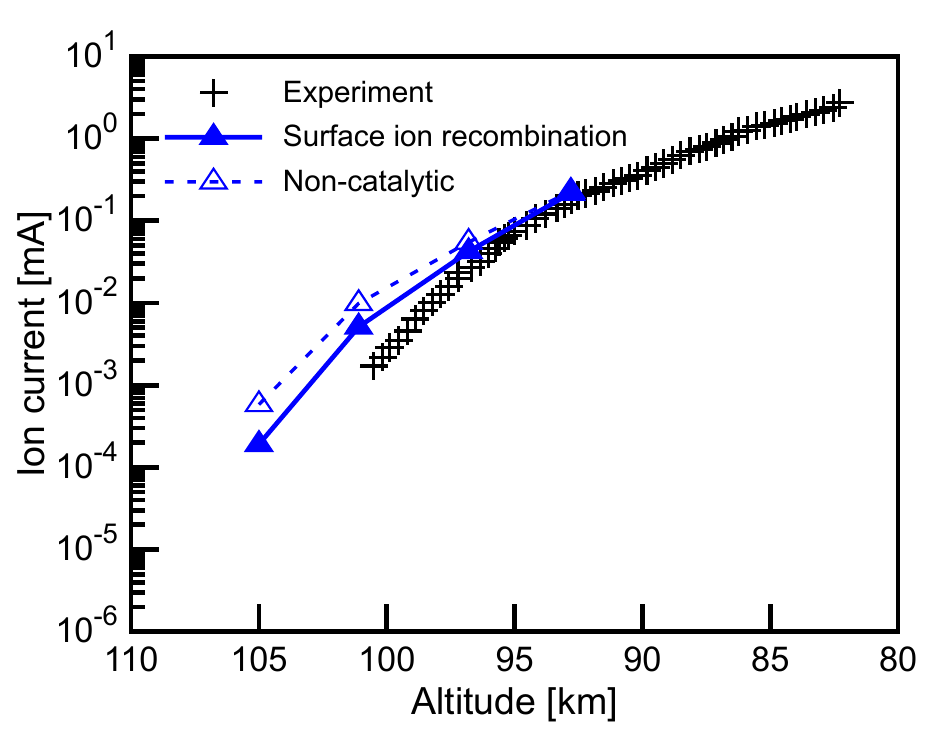}
  \caption{Ion saturation current at Probe 5, $x = 895.33\,\mathrm{mm}, y = 1649.10\,\mathrm{mm}$, at each altitude. The crosses represent the flight measurements. The triangles indicate the present results with the baseline rates listed in Table~\ref{tab:rate_parameters_mole}. Solid lines with filled symbols denote the results with surface ion recombination, and dashed lines with open symbols denote the results with the non-catalytic wall.}
  \label{fig:ion_current}
\end{figure*}

Figure~\ref{fig:ion_current} presents the comparison between the measured ion saturation current at Probe 5, extracted from the plot in Ref.~\cite{orex_1994}, and the values derived from the present simulations with the baseline rates listed in Table~\ref{tab:rate_parameters_mole} for both surface treatments. With surface ion recombination, the computed current was approximately $1.6$ and $1.2$ times the nearest measurements at 96.8 and 92.8~km, respectively, while at 101.1~km it was approximately $3.0$ times the nearest available data point at 100.5~km, indicating an over-prediction at the highest altitude with available data. With the non-catalytic wall, the computed current was approximately $5.8$, $2.0$, and $1.3$ times the measurements at the three altitudes.

The simulation condition at 105.0~km lies outside the range of available experimental data, and the computed currents are therefore predictions. The present SWS enables the estimation of the electron number density at such high altitudes, and the results are consistent with the overall trend observed across the other altitude cases.

As the vehicle descends, the effect of the surface ion recombination becomes weaker. It reduces the Probe-5 electron number density to approximately $0.31$, $0.51$, $0.77$, and $0.96$ times the non-catalytic value at 105.0, 101.1, 96.8, and 92.8~km, respectively. At high altitude, more electrons recombine at the surface than in the gas phase, whereas at lower altitudes the gas-phase recombination becomes dominant and the results of the two surface treatments match. This trend is consistent with the finding of Parent et al.~\cite{parent_electron_2022} that electron loss to the surface dominates over gas-phase recombination at low dynamic pressure.

It should be noted that the correction of the collision probability for the SWS, as expressed in Eq.~\eqref{eq:DSMC_collision_probability_maxfilter}, is essential for accurate prediction. Without this modification, the number of collision attempts between major species particles (such as N and O atoms) is relatively reduced because the selection process becomes biased toward collisions involving low-weighted trace species particles. Since associative ionization reactions are undersampled, the lack of these major-species collision pairs leads to a significant underestimation of the electron number density. The modified collision handling ensures that the combination of accepted collision pairs correctly reflects the physical reaction rates, particularly for these critical electron-producing processes.

One of the possible causes of the increase in the over-prediction with altitude is the reaction model in DSMC. The TCE model converts the Arrhenius rates defined at thermal equilibrium into reaction probabilities based on the total collision energy, in which every energy mode contributes equally~\cite{bird_molecular_1994}, whereas dissociation has been reported to proceed preferentially from vibrationally excited molecules~\cite{haas_models_1993, aiken_review_2025}. The dissociation of $\mathrm{N_2}$ and $\mathrm{O_2}$, which supplies the N and O atoms for associative ionization, can therefore be overestimated where the vibrational temperature lies far below the translational temperature, and this thermal nonequilibrium becomes stronger at higher altitudes. For the RAM-C II flight experiment at altitudes of 73 to 81~km, the effect of the vibrational favoring of dissociation on the electron number density has been investigated~\cite{shevyrin_validation_2014}. The electron number density in the shock layer computed with the TCE model was 10-fold higher than that with a vibrationally favored dissociation model at 81~km, and the difference decreased with decreasing altitude. Investigating this effect requires dissociation models that account for the preferential role of vibrational excitation and is beyond the scope of the present study.

Another possible cause is the uncertainty of the rate coefficients. This is assessed in the following subsection.


\subsection{Sensitivity to reaction rate uncertainty}
\label{sec:rate_uncertainty}
\begin{figure*}[hbtp]
  \centering
  \includegraphics[width=0.55\textwidth]{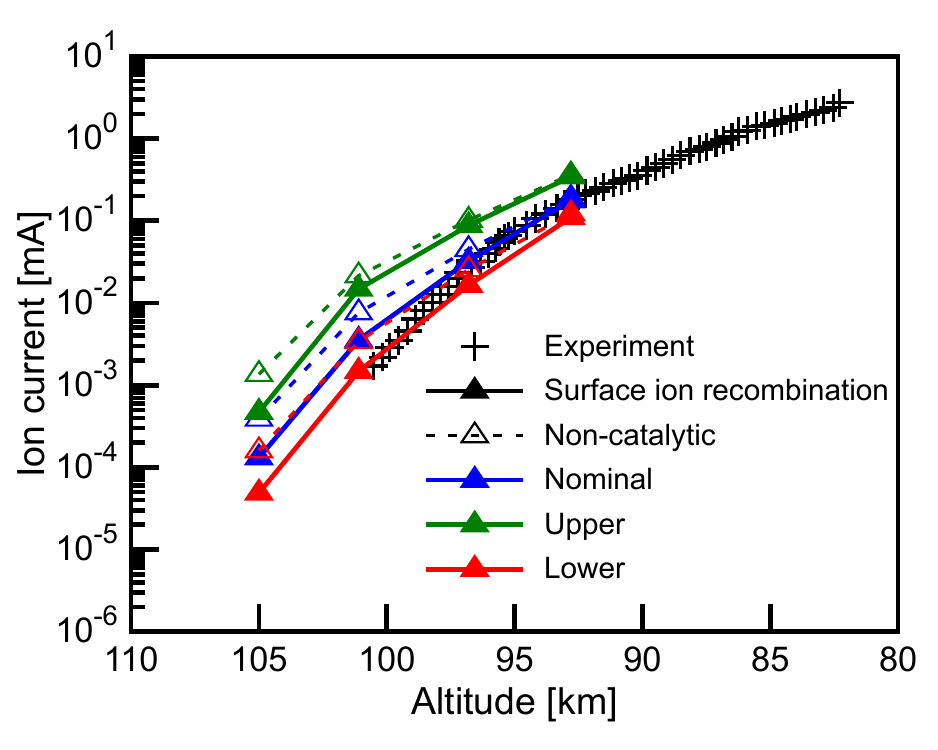}
  \caption{Ion saturation current at Probe 5 with the associative-ionization rates replaced by the values listed in Table~\ref{tab:aiken_rates}. The blue lines show the recommended (nominal) set, and the green and red lines show the results at the upper and lower bounds of the recommended uncertainty interval for the $\mathrm{N}+\mathrm{O}$ associative-ionization rate. Solid lines with filled symbols denote the results with surface ion recombination, and dashed lines with open symbols denote the results with the non-catalytic wall. The crosses represent the flight measurements.}
  \label{fig:ion_current_Aiken}
\end{figure*}

Finally, the range that the computed current can take under the uncertainty of the associative-ionization rates is assessed, as shown in Fig.~\ref{fig:ion_current_Aiken}. As described in Sec.~\ref{sec:case_model}, the forward associative-ionization rates were replaced by the values recommended by Aiken et al.~\cite{aiken_review_2025}, and the $\mathrm{N}+\mathrm{O}\rightarrow\mathrm{NO^+}+\mathrm{e^-}$ rate, the dominant electron source at the present conditions, was varied between the lower and upper bounds of its recommended uncertainty interval. Each rate set was combined with both surface treatments. With surface ion recombination, the computed current ranged from approximately $0.9$ to $8.6$ times the nearest measurement at 101.1~km, from $0.6$ to $3.2$ times at 96.8~km, and from $0.6$ to $1.9$ times at 92.8~km. This range encompasses the available flight measurements. When the non-catalytic wall is also included, the computed current ranged from $0.9$ to $12.5$ times at 101.1~km. The over-prediction of the nominal rates at 101.1~km therefore lies within the rate uncertainty, with the measured current falling between the lower-bound and the nominal rate sets. At 101.1~km, every combination of the rate sets and the surface treatments produced a plasma layer exceeding $n_{\mathrm{e,crit}}$ ahead of the nose. At 105.0~km, the electron number density obtained with every rate set with surface ion recombination remained below $n_{\mathrm{e,crit}}$ on the stagnation streamline, and that obtained with the non-catalytic wall exceeded it only near the forebody surface. Therefore, the prediction of the VHF blackout onset at approximately 101.1~km or thereafter holds under both surface treatments and the rate uncertainty.

\clearpage

\section{CONCLUSION}
\label{sec:conclusion}

This paper extended the SWS within the DSMC framework to chemically reactive flows. 
This method assigns species-dependent numerical weights $w_i$ and combines a split-merge procedure for momentum and energy conservation with particle production and the probabilistic treatment of particle creation and deletion in reactive collisions when reactant and product species weights are unequal.
To correct the selection bias caused by the standard NTC algorithm when simulated particle counts do not correspond to the actual mixture composition, a modified collision-pair acceptance probability was used, ensuring that the electron-producing associative ionization reactions occur at the correct physical rates.
The implementation was first assessed in two single-cell test cases involving a dissociation reaction and an exchange reaction with species weight ratios of ten. 
The results confirmed that total mass is conserved exactly when weight ratios are integers, and conserved on average otherwise. 
The total particle energy of the system was also found to be conserved on average in the presence of the chemical reactions. The mole-fraction histories agreed with those from conventional DSMC until the weighted species became a major component of the mixture.
The validated reactive SWS was applied to the OREX flight reentry cases, consisting of the four conditions spanning altitudes from 105.0~km to 92.8~km. Translational and internal energy mode temperatures of the species mixture agreed closely with those from conventional DSMC, confirming that the trace species simulation using the species weights does not produce a deviation in the macroscopic flow field. The SWS additionally yields spatially continuous electron number density distributions that are inaccessible with conventional DSMC at these altitudes. 
The simulated ion saturation currents at the electrostatic probe location reproduced the overall trend of the available flight measurements and agreed with them at $96.8$ and $92.8$~km, while the computed current at $101.1$~km was approximately $3.0$ times the nearest measurement. The range of the computed currents under the uncertainty of the associative-ionization rates and the two treatments of the surface ion recombination encompassed these measurements.
The spatial extent of the critical electron number density isoline, based on the electron number density field, suggests VHF blackout onset at approximately 101.1~km or thereafter, consistent with the flight observation.
This prediction held under both surface treatments and the reaction-rate uncertainty.
By maintaining species-assigned weights throughout the simulation, the present approach keeps the computational load manageable. The resulting smooth electron number density fields are directly suitable as inputs for electromagnetic wave propagation analyses, providing a practical computational method for blackout prediction and the evaluation of mitigation strategies in nonequilibrium rarefied ionized hypersonic flows. The present framework is further applicable to other reactive rarefied flow scenarios where accurate trace species solutions are essential, such as combustion in micro-channel flows, chemical vapor deposition and atomic layer deposition processes, and plume/spacecraft interaction and associated contamination.

\begin{acknowledgments}
This work was supported by JSPS KAKENHI (Grant Numbers JP24KF0039, JP25KJ1404), Sumitomo Electric Group CSR Foundation, and Foundation of Public Interest of Tatematsu. Part of the computer simulation was performed on the KDK computer system at the Research Institute for Sustainable Humanosphere, Kyoto University. The authors thank Wataru Sarae of Japan Aerospace Exploration Agency for valuable discussions.
\end{acknowledgments}

\section*{Author Declarations}
\subsection*{Conflict of Interest}

The authors have no conflicts to disclose.

\subsection*{Data Availability}
The data that support the findings of this study are available from the corresponding author upon reasonable request.

\newpage
\appendix
\section{Convergence and sensitivity of the numerical parameters}
\label{sec:appendix_convergence}

This appendix assesses the sensitivity of the computed fields to the three numerical parameters, the global weighting factor $F_N$, the species weights $w_i$, and the averaging duration, in order to confirm that the values adopted in the present study are appropriate to obtain converged fields. Each parameter was halved and doubled about its baseline value in Table~\ref{tab:computational_conditions} at all four altitudes, one parameter at a time, with the remaining parameters held fixed. The assessment was performed with the non-catalytic wall condition, since these parameters affect only the statistical behavior.

\begin{figure*}[hbtp]
  \centering
  \includegraphics[width=0.95\textwidth]{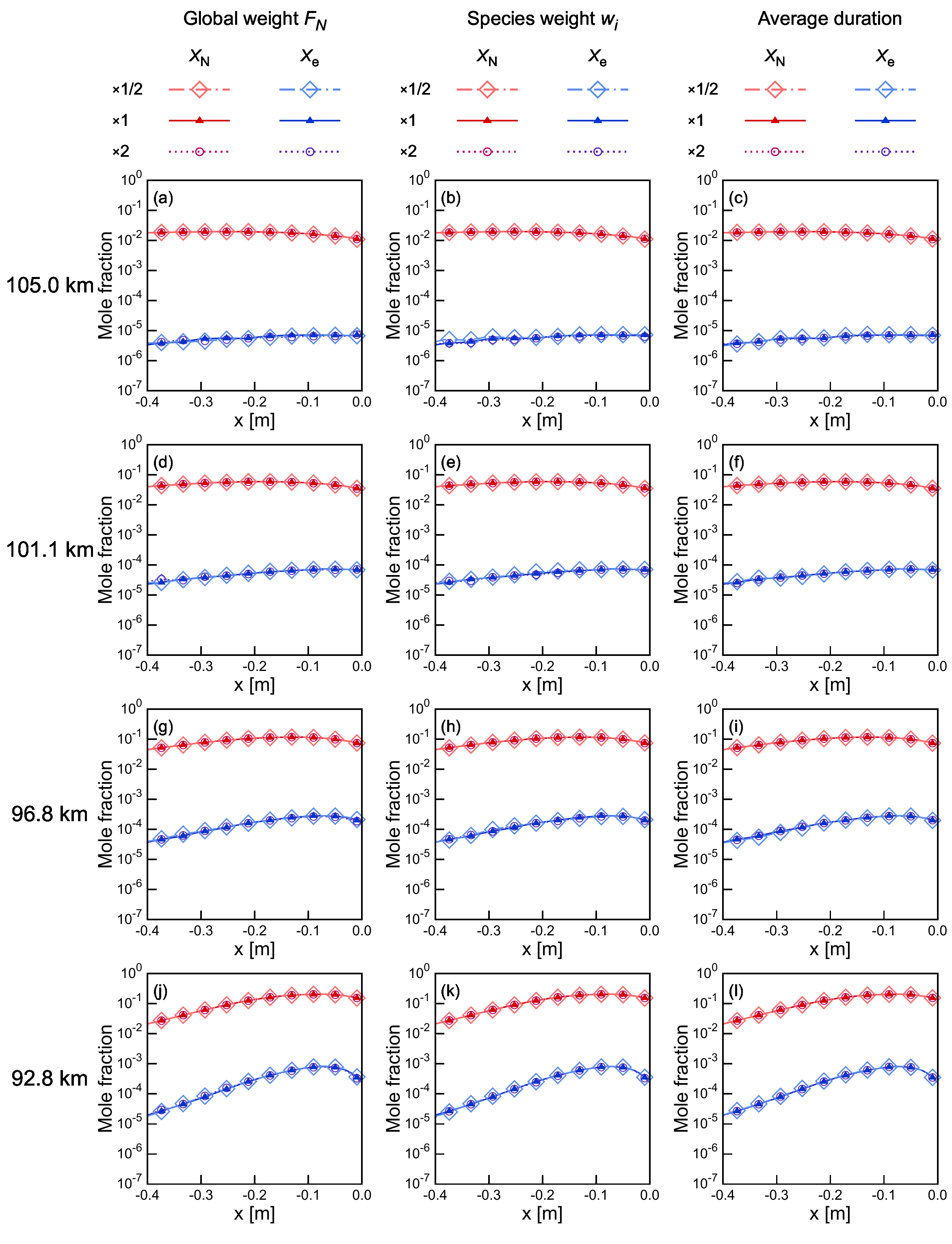}
  \caption{Mole fractions of atomic nitrogen $X_\mathrm{N}$ and electrons $X_\mathrm{e}$ along the stagnation streamline. Each column corresponds to one parameter that was halved and doubled about its baseline value, the global weighting factor $F_N$, the species weight $w_i$, and the averaging duration, and each row to one altitude (105.0, 101.1, 96.8, and 92.8 km). The individual graphs are labeled (a)--(l).}
  \label{fig:sensitivity-mole-fraction}
\end{figure*}

Figure~\ref{fig:sensitivity-mole-fraction} compares the stagnation-streamline mole fractions of atomic nitrogen $X_\mathrm{N}$ and electrons $X_\mathrm{e}$ obtained using the halved and doubled values of each parameter against the baseline profiles. Atomic nitrogen was chosen as a major product of the shock layer chemistry and a reactant of the associative ionization that produces the electrons, and the electron mole fraction represents the trace species of interest. 

The global weighting factor $F_N$ affects the total number of simulated particles, and a smaller value gives a larger number of particles and hence better converged statistics. In the graphs of the global weighting factor $F_N$, namely (a), (d), (g), and (j), a small deviation from the baseline profile was observed when $F_N$ was doubled, whereas the halved value agreed closely with the baseline. The species weight $w_i$ of the charged species governs the number of simulated electrons in particular, and a smaller weight improves the statistical convergence of the electron mole fraction. It does not affect the physical result, as the number of real molecules represented by the simulated particles is the same. The profiles obtained with the baseline and with the halved $w_i$ were almost indistinguishable. The averaging duration determines the smoothness of the time-averaged fields, and a longer duration gives a smoother profile. In the graphs (c), (f), (i), and (l), the mole fraction distribution was resolved even with the halved averaging duration, which shows that the duration adopted in the present study is sufficient.
The atomic nitrogen profiles were unchanged for all three parameters at every altitude. The electron mole fraction exhibited larger fluctuations at $105.0$~km, where the number of simulated electrons is smallest, whereas the atomic nitrogen profile at this altitude was also unchanged, which indicates that these fluctuations arise from the sampling of the trace species and not from a change in the flow field. It was therefore confirmed that the values adopted in the present computations are appropriate to obtain converged fields.

\clearpage

\bibliography{reference} 

\clearpage

\end{document}